\documentclass[12pt,reqno,a4paper]{amsart}
\usepackage{amsmath,amssymb,dsfont,graphicx,cite,latexsym,epsf,cancel,
tikz,
subfig
}
\usepackage[colorlinks=false]{hyperref}
\usetikzlibrary{arrows}
\newtheorem{theorem}{Theorem}

\makeatletter
\expandafter\let\expandafter
\reset@font\csname reset@font\endcsname
\def\subeqnarray{\arraycolsep1pt
   \def\@eqnnum\stepcounter##1{\stepcounter{subequation}
       {\reset@font\rm(\theequation\alph{subequation})}}
\jot5mm     \eqnarray}

\makeatother

\newcommand{\bbZ}{{\mathbb Z}}
\newcommand{\bbR}{{\mathbb R}}
\newcommand{\cL}{{\mathcal L}}

\def\epsilon{\varepsilon}

\def\tilde{\widetilde}

\def\endpf{$\blacksquare$\medskip}
\def\wx{\widetilde{x}}
\def\wip{\widetilde{p}}
\def\whx{\widehat{x}}
\def\whp{\widehat{p}}

\newbox\meibox
\def\placeunder#1#2#3#4{\setbox\meibox%
\vbox{\hbox{\hskip#4$\hphantom{#2}$}\hbox{$\hphantom{#1}$}}%
\vtop{\baselineskip=0pt\lineskiplimit=\baselineskip%
\lineskip=#3\hbox to \wd\meibox{\hfil\hskip#4$#2$\hfil}%
\hbox to \wd\meibox{\hfil$#1$\hfil}}}
\def\undertilde#1{\mathchoice{%
\placeunder{\vbox to 1.4pt{\hbox{$\displaystyle\widetilde{\,\,\,
}$}\vss}}{\displaystyle#1}{1.5pt}{1.5pt}}%
{\placeunder{\vbox to 1.4pt{\hbox{$\textstyle\widetilde{\,\,
}$}\vss}}{\textstyle#1}{1.5pt}{1.5pt}}%
{\placeunder{\vbox to 1.4pt{\hbox{$\scriptstyle\tilde{
}$}\vss}}{\scriptstyle#1}{1pt}{1pt}}%
{\placeunder{\vbox to 1.4pt{\hbox{$\scriptscriptstyle\tilde{
}$}\vss}}{\scriptscriptstyle#1}{1pt}{1pt}}%
}
\def\intprod{\mathbin{\hbox to 6pt{%
                 \vrule height0.4pt width5pt depth0pt
                 \kern-.4pt
                 \vrule height6pt width0.4pt depth0pt\hss}}}
\begin{document}
\title[Multi-time Lagrangian 1-forms]
{Multi-time Lagrangian 1-forms \\ for families of B\"acklund transformations.\\
Toda-type systems}

\author{Raphael Boll, Matteo Petrera, Yuri B. Suris }

\maketitle

\begin{center}
{\footnotesize{
Institut f\"ur Mathematik, MA 7-2,\\
Technische Universit\"at Berlin, Str. des 17. Juni 136\\
10623 Berlin, Germany\\
E-mail: {\tt boll, petrera, suris@math.tu-berlin.de}
}}
\end{center}


\begin{abstract} General Lagrangian theory of discrete one-dimensional integrable systems is illustrated by a detailed study of B\"acklund transformations for Toda-type systems. Commutativity of B\"acklund transformations is shown to be equivalent to consistency of the system of discrete multi-time Euler-Lagrange equations. The precise meaning of the commutativity in the periodic case, when all maps are double-valued, is established. It is shown that gluing of different branches is governed by the so called superposition formulas. The closure relation for the multi-time Lagrangian 1-form on solutions of the variational equations is proved for all Toda-type systems. Superposition formulas are instrumental for this proof. The closure relation was previously shown to be equivalent to the spectrality property of B\"acklund transformations, i.e., to the fact that the derivative of the Lagrangian with respect to the spectral parameter is a common integral of motion of the family of B\"acklund transformations. We relate this integral of motion to the monodromy matrix of the zero curvature representation which is derived directly from equations of motion in an algorithmic way. This serves as a further evidence in favor of the idea that B\"acklund transformations serve as zero curvature representations for themselves.
\end{abstract}

\section{Introduction}

The present paper can be considered as an extended illustration of the general Lagrangian theory of discrete integrable systems of classical mechanics, developed in \cite{S12}. This development was prompted by an example of the discrete time Calogero-Moser system studied in \cite{YLN}, and belongs to the line of research on variational formulation of more general discrete integrable systems, initiated by Lobb and Nijhoff in \cite{LN1}.

The notion of integrability of discrete systems, lying at the basis of this development, is that of the multidimensional consistency.  This understanding of integrability of discrete systems has been a major breakthrough \cite{BS1}, \cite{N}, and stimulated an impressive activity boost in the area, cf. \cite{DDG}. According to the concept of multi-dimensional consistency, integrable $d$-dimensional systems can be imposed in a consistent way on all $d$-dimensional sublattices of a lattice $\bbZ^m$ of arbitrary dimension. This means that the resulting multi-dimensional system possesses solutions whose restrictions to any $d$-dimensional sublattice are generic solutions of the corresponding two-dimensional system. In the case $d=1$ the concept of multi-dimensional consistency is more or less synonymous with the idea of integrability as commutativity which has been advocated by Veselov \cite{V}. In the important case of discrete integrable systems of dimension $d=2$, this approach led to classification results \cite{ABS} (ABS list) which turned out to be rather influential.

The Lagrangian aspects of the theory were, as mentioned above, pushed forward in \cite{LN1}. They observed that the value of the action functional for ABS equations remains invariant under local changes of the underlying quad-surface, and suggested to consider this as a defining feature of integrability. Their results, found on the case-by-case basis for some equations of the ABS list, have been extended to the whole list and given a more conceptual proof in \cite{BS2}, and have been subsequently generalized in various directions: for multi-field two-dimensional systems \cite{LN2}, \cite{ALN}, for asymmetric two-dimensional systems \cite{BollSuris}, for dKP, the fundamental three-dimensional discrete integrable system \cite{LNQ}, and for the above mentioned example of one-dimensional integrable systems \cite{YLN}.

General Lagrangian theory of discrete one-dimensional integrable systems has been developed in \cite{S12}. We give a short account of this theory in Section \ref{sect: discr results}. It turns out that the most significant class of examples is given by {\em B\"acklund transformations}, i.e., one-parameter families of commuting symplectic maps. In the main body of this paper, Sections \ref{sect: BT Toda}--\ref{sect: BT mult hyp}, we work out the relevant results for B\"acklund transformations for integrable systems of the Toda type. The list of systems under consideration is given, for convenience of the reader, in Section \ref{sect: Todas}. Our main results are the following.
\begin{enumerate}
\item Although B\"acklund transformations for the Toda lattice are very well studied (cf. \cite{KS} and quotations therein), there is one aspect which was not sufficiently dealt with in the existing literature. Usually one considers these systems under periodic boundary conditions, which yields {\em multi-valuedness} of the  corresponding maps. A general discussion of commutativity of multi-valued maps (correspondences) is contained in \cite{V}. However, its applicability to B\"acklund transformations of Toda-like systems, in particular, the choice of branches ensuring commutativity of such maps seems to be a completely open problem. We give here a complete solution of this problem. The key ingredient are the so called {\em superposition formulas}, which enable us to precisely describe the branching behavior on the level of local algebraic relations.
\item The main feature of the Lagrangian theory of discrete integrable systems is the so called {\em closure relation}, which expresses the fact that the Lagrangian form on the multi-dimensional space of independent variables is closed on solutions of variational equations. In the case of B\"acklund transformations, it was shown in \cite{S12} that the closure relation is equivalent to the so called {\em spectrality property} introduced by Sklyanin and Kuznetsov in \cite{KS}, which had, up to now, a somewhat mysterious reputation. We prove spectrality (and thus the closure relation) for all systems of the Toda type. Superposition formulas turn out to be of a crucial importance also for this aim.
\item In \cite{S12}, an idea was pushed forward that Lax representations for B\"acklund transformations are already encoded in the equations of motion themselves. This is a one-dimensional counterpart of an analogous idea for two-dimensional systems, which was one of the main breakthroughs of \cite{BS1}, \cite{N}. Here, we support this idea by an algorithmic derivation of transition and monodromy matrices for all Toda-type integrable systems. We think that a great portion of a mystic flair still enjoyed by integrable systems gets herewith a rational and ultimately simple explanation.
\end{enumerate}

\section{General theory of discrete multi-time Euler-Lagrange equations}
\label{sect: discr results}

\begin{figure}[tbp]
\subfloat[]{
\begin{tikzpicture}[auto,scale=1.5,>=stealth',inner sep=3pt]
   \node (x) at (0,0) [circle,fill,thick,label=-135:$x$,label=45:$\left(E\right)$] {};
   \node (x1) at (2,0) [circle,fill,thick,label=135:$\left(E_{1}\right)$,label=-45:$\wx$] {};
   \node (x2) at (0,2) [circle,fill,thick,label=-45:$\left(E_{2}\right)$,label=135:$\whx$] {};
   \node (x12) at (2,2) [circle,fill,thick,label=45:$\widehat{\wx}$,label=-135:$\left(E_{12}\right)$] {};
   \draw [thick,->] (x) to (x1) to (x12);
   \draw [thick,->] (x) to (x2) to (x12);
\end{tikzpicture}
}\qquad
\subfloat[]{
\begin{tikzpicture}[auto,scale=1.5,>=stealth',inner sep=3pt]
   \node (x) at (0,0) [circle,fill,thick,{label=-135:$\left(x,p\right)$}] {};
   \node (x1) at (2,0) [circle,fill,thick,{label=-45:$\left(\wx,\wip\right)$}] {};
   \node (x2) at (0,2) [circle,fill,thick,{label=135:$\left(\whx,\whp\right)$}] {};
   \node (x12) at (2,2) [circle,fill,thick,{label=45:$\big(\widehat{\wx},\widehat{\wip}\big)$}] {};
   \draw [thick,->] (x) to node {$F_{\lambda}$} (x1);
   \draw [thick,->] (x) to node [swap] {$F_{\mu}$} (x2);
   \draw [thick,->] (x2) to node [swap]{$F_{\lambda}$} (x12);
   \draw [thick,->] (x1) to node {$F_{\mu}$} (x12);
\end{tikzpicture}
}
\caption{Consistency of multi-time Euler-Lagrange equations: (a) Start with data $x$, $\wx$, $\whx$ related by corner equation (\ref{eq: E}); solve corner equations (\ref{eq: E1}) and (\ref{eq: E2}) for $\widehat{\wx}$; consistency means that the two values of $\widehat{\wx}$ coincide identically and satisfy corner equation (\ref{eq: E12}). (b) Maps $F_{\lambda}$ and $F_{\mu}$ commute.}
\label{Fig: consistency}
\end{figure}
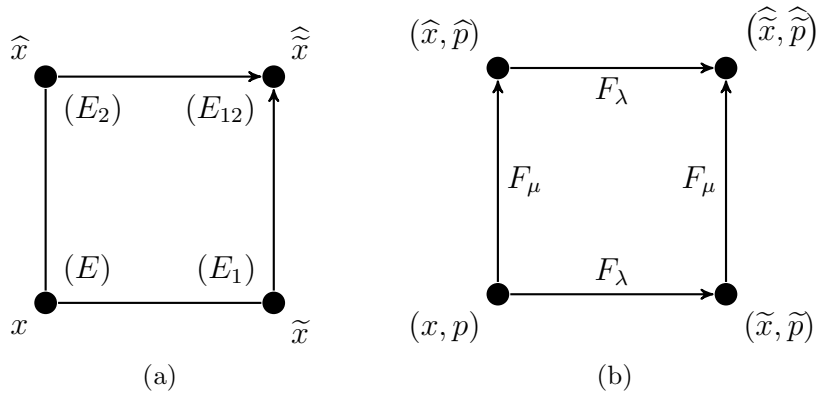

We will now recall the main positions of the Lagrangian theory of discrete one-dimen\-sional integrable systems, developed in \cite{S12}, in application to {\em B\"acklund transformations}, i.e., to one-parameter families of commuting symplectic maps.  Suppose that such a symplectic map $F_\lambda:(x,p)\mapsto(\widetilde x,\widetilde p)$, depending on the parameter $\lambda$, admits a generating function $\Lambda$:
\begin{equation}\label{eq: BT1}
F_\lambda:\
p=-\frac{\partial \Lambda(x,\wx;\lambda)}{\partial x},\quad
\wip=\frac{\partial \Lambda(x,\wx;\lambda)}{\partial \wx}.
\end{equation}
Here, the first equation should be (at least locally) solvable for $\widetilde x$, i.e.,
the matrices of the mixed second order partial derivatives of the Lagrange function $\Lambda$ should be non-degenerate, $\det(\partial^2 \Lambda/\partial x\partial\widetilde x)\neq 0$.
When considering a second such map, say $F_\mu$, we will denote its action by a hat:
\begin{equation}\label{eq: BT2}
F_\mu:\
p=-\frac{\partial \Lambda(x,\whx;\mu)}{\partial x},\quad
\whp=\frac{\partial \Lambda(x,\whx;\mu)}{\partial \whx}.
\end{equation}
We assume that
\begin{equation*}\label{eq: commute}
F_\lambda\circ F_\mu=F_\mu\circ F_\lambda.
\end{equation*}
As a consequence, the following equations, called corner equations, are obtained by eliminating $p$ from (\ref{eq: BT1}), (\ref{eq: BT2}) at the four vertices of a square on Fig. \ref{Fig: consistency}:
\begin{equation}\label{eq: E}\tag{$E$}
\frac{\partial\Lambda(x,\wx;\lambda)}{\partial x}-
\frac{\partial\Lambda(x,\whx;\mu)}{\partial x}=0,
\end{equation}
\begin{equation}\label{eq: E1}\tag{$E_1$}
\frac{\partial\Lambda(x,\wx;\lambda)}{\partial \wx}+
\frac{\partial\Lambda(\wx,\widehat{\wx};\mu)}{\partial \wx}=0,
\end{equation}
\begin{equation}\label{eq: E2}\tag{$E_2$}
\frac{\partial\Lambda(x,\whx;\mu)}{\partial \whx}+
\frac{\partial\Lambda(\whx,\widehat{\wx};\lambda)}{\partial \whx}=0,
\end{equation}
and
\begin{equation}\label{eq: E12}\tag{$E_{12}$}
\frac{\partial\Lambda(\whx,\widehat{\wx};\lambda)}{\partial \widehat{\wx}}-
\frac{\partial\Lambda(\wx,\widehat{\wx};\mu)}{\partial \widehat{\wx}} = 0.
\end{equation}
These equations admit the following variational interpretation. We define the {\em discrete multi-time Lagrangian 1-form} for a family of B\"acklund transformations as a discrete 1-form whose values on the (directed) edges of $\mathbb Z^2$ are given by $\Lambda(x,\wx;\lambda)$, resp. $\Lambda(x,\whx;\mu)$. A generalization to $\mathbb Z^m$ with any $m\ge 2$ is straightforward. We look for functions $x:\mathbb Z^m\to\mathbb R$ delivering critical points for the action along {\em any} discrete curve in $\mathbb Z^m$. Then equations (\ref{eq: E})--(\ref{eq: E12}) are nothing but {\em multi-time Euler-Lagrange equations} for this variational problem; see \cite{S12}.

Consistency of the system of multi-time Euler-Lagrange equations (\ref{eq: E})--(\ref{eq: E12}) should be understood as follows: start with the fields $x$, $\wx$, $\whx$ satisfying the corner equation (\ref{eq: E}). Then each of the corner equations (\ref{eq: E1}), (\ref{eq: E2}) can be solved for $\widehat{\wx}$. Thus, we obtain two alternative values for the latter field. Consistency takes place if these values coincide identically (with respect to the initial data), and, moreover, if the resulting field $\widehat{\wx}$ satisfies the corner equation (\ref{eq: E12}). This is equivalent to commutativity of $F_\lambda$ and $F_\mu$. See Fig. \ref{Fig: consistency}.

We mention that the standard single-time Euler-Lagrange equations for the maps $F_\lambda$,
\begin{equation*}\label{eq: dEL}
\frac{\partial\Lambda(\undertilde{x},x;\lambda)}{\partial x}+
\frac{\partial\Lambda(x,\wx;\lambda)}{\partial x}=0,
\end{equation*}
are a consequence of equation (\ref{eq: E}) and (the downshifted version of) equation (\ref{eq: E1}).

As shown in \cite{S12}, on solutions of discrete multi-time Euler-Lagrange equations, we have:
\begin{equation}\label{eq: BT closure}
    \Lambda(x,\wx;\lambda)+\Lambda(\wx,\widehat{\wx};\mu)-
    \Lambda(x,\whx;\mu)-\Lambda(\whx,\widehat{\wx};\lambda)=\ell(\lambda,\mu)={\rm const}.
\end{equation}
Moreover, $\ell(\lambda,\mu)=0$, i.e., the discrete multi-time Lagrangian 1-form is closed on solutions, if and only if $\partial \Lambda(x,\wx;\lambda)/\partial \lambda$ is a common integral of motion for all $F_\mu$. The latter property is a re-formulation of the mysterious ``spectrality property'' of B\"acklund transformations discovered by Kuznetsov and Sklyanin \cite{KS}.

\section{Toda-type systems and their time discretizations}
\label{sect: Todas}

We will illustrate the above concepts with an important and representative set of examples, namely, B\"acklund transformations for Toda-type systems. The latter term is used to denote integrable lattice equations of the general form
\begin{equation*}\label{eq: Toda type}
\ddot{x}_k=r(\dot{x}_k)\big(f(x_{k+1}-x_k)-f(x_k-x_{k-1})\big).
\end{equation*}
The integrable discretizations \cite{S} are of the form
\begin{equation*}\label{eq: dToda type}
g(\wx_k-x_k;h)-g(x_k-\undertilde{x}_k;h)=f(\undertilde{x}_{k+1}-x_k;h)-f(x_k-\wx_{k-1};h),
\end{equation*}
with $h$ being an arbitrary parameter (time step). It is this parameter (or its inverse) which will play the role of the B\"acklund parameter $\lambda$ in all our examples. The list of examples includes:
\begin{itemize} 
\item The original (exponential) {\em Toda lattice:}
\begin{equation}\label{eq: Toda}
\ddot{x}_k=e^{x_{k+1}-x_k}-e^{x_k-x_{k-1}},
\end{equation}
with a discrete time counterpart
\begin{equation}\label{eq: d Toda}
e^{\wx_k-x_k}-e^{x_k-\undertilde{x}_k}=
h^2\left(e^{\undertilde{x}_{k+1}-x_k}-e^{x_k-\wx_{k-1}}\right).
\end{equation}

\item {\em Dual Toda lattice:}
\begin{equation}\label{eq: dual Toda}
\ddot{x}_k=\dot{x}_k\left(x_{k+1}-2x_k+x_{k-1}\right),
\end{equation}
with a discrete time counterpart
\begin{equation}\label{eq: d dual Toda}
\frac{\wx_k-x_k}{x_k-\undertilde{x}_k}=
\frac{\,\raisebox{0.5mm}{$1+h(\undertilde{x}_{k+1}-x_k)$}}
{\,\raisebox{-0.5mm}{$1+h(x_k-\wx_{k-1})$}}.
\end{equation}

\item {\em Modified Toda lattice:}
\begin{equation}\label{eq: mod Toda}
\ddot{x}_k=\dot{x}_k\big(e^{x_{k+1}-x_k}-e^{x_k-x_{k-1}}\big),
\end{equation}
with a discrete time counterpart
\begin{equation}\label{eq: d mod Toda}
\frac{e^{\wx_k-x_k}-1}{e^{x_k-\undertilde{x}_k}-1}=
\frac{1+he^{\undertilde{x}_{k+1}-x_k}}{1+he^{x_k-\wx_{k-1}}}\;.
\end{equation}

\item {\em Symmetric rational additive Toda-type system:}
\begin{equation}\label{eq: sym rat add Toda}
\ddot{x}_k=-\dot{x}_k^2\left(\frac{1}{x_{k+1}-x_k}-\frac{1}{x_k-x_{k-1}}\right),
\end{equation}
with a discrete time counterpart
\begin{equation}\label{eq: d sym rat add Toda}
\frac{1}{\wx_k-x_k}-\frac{1}{x_k-\undertilde{x}_k}=
\frac{1}{\undertilde{x}_{k+1}-x_k}-\frac{1}{x_k-\wx_{k-1}}.
\end{equation}

\item {\em Symmetric rational multiplicative Toda-type system:}
\begin{equation}\label{eq: sym rat mult Toda}
\ddot{x}_k=-(\dot{x}_k^2-1)\left(\frac{1}{x_{k+1}-x_k}-\frac{1}{x_k-x_{k-1}}\right),
\end{equation}
with a discrete time counterpart
\begin{equation}\label{eq: d sym rat mult Toda}
\frac{(\wx_k-x_k+h)}{(\wx_k-x_k-h)}\cdot
\frac{(x_k-\undertilde{x}_k-h)}{(x_k-\undertilde{x}_k+h)}
=\frac{(\undertilde{x}_{k+1}-x_k+h)}{(\undertilde{x}_{k+1}-x_k-h)}
\cdot\frac{(x_k-\wx_{k-1}-h)}{(x_k-\wx_{k-1}+h)}.
\end{equation}

\item {\em Symmetric hyperbolic multiplicative Toda-type system:}
\begin{equation}\label{eq: sym hyp mult Toda}
\ddot{x}_k=-(\dot{x}_k^2-1)\big(\coth(x_{k+1}-x_k)-\coth(x_k-x_{k-1})\big),
\end{equation}
with a discrete time counterpart
\begin{equation}\label{eq: d sym hyp mult Toda}
\frac{\sinh(\wx_k-x_k+h)}{\sinh(\wx_k-x_k-h)}\cdot
\frac{\sinh(x_k-\undertilde{x}_k-h)}{\sinh(x_k-\undertilde{x}_k+h)}
=\frac{\sinh(\undertilde{x}_{k+1}-x_k+h)}{\sinh(\undertilde{x}_{k+1}-x_k-h)}
\cdot\frac{\sinh(x_k-\wx_{k-1}-h)}{\sinh(x_k-\wx_{k-1}+h)}\;.
\end{equation}
\end{itemize}
(The names of the last three systems are justified by the appearance of their discrete time versions.)

We consider these systems with finitely many degrees of freedom $(1\le k\le N)$. This requires to specify certain boundary conditions. We will consider either periodic boundary conditions (all indices taken ${\rm mod}\, N$, so that $x_0=x_N$, $x_{N+1}=x_1$), or the so called open-end boundary conditions, which can be imposed in all cases except for the dual Toda lattice and which can be formally achieved by setting $x_0=\infty$ and $x_{N+1}=-\infty$. For the dual Toda lattice (and for the modified Toda lattice, as well), a certain ersatz for the open-end boundary conditions exists. It is achieved by considering the periodic system with $N+1$ particles enumerated by $0\le k\le N$ and by restricting it to $x_0=x_{N+1}=0$. For such a reduction, all results of the present paper can be established, but we do not deal with it in detail, since the resulting systems have a somewhat different flavor of the affine root
system $C_N^{(1)}$ rather than the classical root system $A_{N-1}$. In particular, these systems do not depend on differences $x_{k+1}-x_k$ alone, due to the presence of the boundary terms depending on $x_1$ and on $x_N$, and therefore the total momentum $\sum_{k=1}^N p_k$ is not a conserved quantity for them.

\section{B\"acklund transformations for Toda lattice}
\label{sect: BT Toda}

Here we illustrate the main constructions by the well-known example of B\"acklund transformations for the Toda lattice (\ref{eq: Toda}). The maps $F_\lambda:\bbR^{2N}\to\bbR^{2N}$ are given by equations of the type (\ref{eq: BT1}):
\begin{equation}\label{eq: BT1 Toda}
F_\lambda:\ \left\{\begin{array}{l}
p_k=\dfrac{1}{\lambda}\left(e^{\wx_k-x_k}-1\right)+\lambda e^{x_k-\wx_{k-1}},\vspace{0.2truecm}\\
\wip_k=\dfrac{1}{\lambda}\left(e^{\wx_k-x_k}-1\right)+\lambda e^{x_{k+1}-\wx_k},
\end{array}\right.
\end{equation}
cf. \cite{TW}, \cite{KS}, \cite{S}.
The corresponding Lagrangian is given by
\begin{equation}\label{eq: BT Toda Lagr}
\Lambda(x,\wx;\lambda)=\frac{1}{\lambda}\sum_{k=1}^N\left(e^{\wx_k-x_k}-1-(\wx_k-x_k)\right)-
    \lambda\sum_{k=1}^N e^{x_{k+1}-\wx_k},
\end{equation}
and the standard single-time Euler-Lagrange equations coincide with (\ref{eq: d Toda}) with $h=\lambda$.
 
In the open-end case, we omit the term with $e^{x_1-\wx_0}$ from the expression for $p_1$, the term with $e^{x_{N+1}-\wx_N}$ from the expression for $\wip_N$, and we let the second sum in (\ref{eq: BT Toda Lagr}) extend over $1\le k\le N-1$ only. In this case, the first equations in (\ref{eq: BT1 Toda}) are uniquely solved for $\wx_1$, $\wx_2$, $\ldots$, $\wx_N$ (in this order), to give
\[
e^{\wx_1-x_1}=1+\lambda p_1,\quad
e^{\wx_2-x_2}=1+\lambda p_2-\frac{\lambda^2e^{x_2-x_1}}{1+\lambda p_1},\quad\cdots\quad,
\]
\begin{equation*}\label{eq: BT Toda expl}
e^{\wx_N-x_N}=1+\lambda p_N-\frac{\lambda^2e^{x_N-x_{N-1}}}{1+\lambda p_{N-1}-
\dfrac{\lambda^2e^{x_{N-1}-x_{N-2}}}{1+\lambda p_{N-2}-\;
\raisebox{-3mm}{$\ddots$}
\raisebox{-4.5mm}{$\;-\dfrac{\lambda^2 e^{x_2-x_1}}{1+\lambda p_1}$}}}.
\end{equation*}
In the periodic case, all $e^{\wx_k-x_k}$ can be expressed as analogous infinite periodic continued fractions, and are, therefore, {\it double-valued} functions of $(x,p)$.

As discussed in the previous section, commutativity of the maps $F_\lambda$, $F_\mu$ (in the open-end case, when they are well-defined, i.e., single-valued) is equivalent to consistency of the system of corner equations:
\begin{equation} \label{eq: BT Toda E}\tag{$E$}
 \dfrac{1}{\lambda}\left(e^{\wx_k-x_k}-1\right)+\lambda e^{x_k-\wx_{k-1}}=
 \dfrac{1}{\mu}\left(e^{\whx_k-x_k}-1\right)+\mu e^{x_k-\whx_{k-1}},
\end{equation}
\begin{equation} \label{eq: BT Toda E1}\tag{$E_1$}
 \dfrac{1}{\lambda}\left(e^{\wx_k-x_k}-1\right)+\lambda e^{x_{k+1}-\wx_k}=
 \dfrac{1}{\mu}\left(e^{\widehat{\widetilde{x}}_k-\wx_k}-1\right)
 +\mu e^{\wx_k-\widehat{\widetilde{x}}_{k-1}},
\end{equation}
\begin{equation} \label{eq: BT Toda E2}\tag{$E_2$}
 \dfrac{1}{\mu}\left(e^{\whx_k-x_k}-1\right)+\mu e^{x_{k+1}-\whx_k}=
 \dfrac{1}{\lambda}\left(e^{\widehat{\widetilde{x}}_k-\whx_k}-1\right)
 +\lambda e^{\whx_k-\widehat{\widetilde{x}}_{k-1}},
\end{equation}
\begin{equation} \label{eq: BT Toda E12}\tag{$E_{12}$}
 \dfrac{1}{\lambda}\left(e^{\widehat{\widetilde{x}}_k-\whx_k}-1\right)
 +\lambda e^{\whx_{k+1}-\widehat{\widetilde{x}}_k}=
 \dfrac{1}{\mu}\left(e^{\widehat{\widetilde{x}}_k-\wx_k}-1\right)
 +\mu e^{\wx_{k+1}-\widehat{\widetilde{x}}_k}.
\end{equation}
We have to clarify the meaning of the both notions (commutativity of  $F_\lambda$, $F_\mu$ and consistency of corner equations) in the periodic case. To do this, we prove the following statement.
\begin{theorem}\label{th: superposition}
Suppose that the fields $x$, $\wx$, $\whx$ satisfy corner equations (\ref{eq: BT Toda E}). Define the fields $\widehat{\widetilde{x}}$ by any of the following two formulas, which are equivalent by virtue of (\ref{eq: BT Toda E}):
\begin{equation}\label{eq: BT Toda S1}\tag{$S1$}
\dfrac{1}{\lambda}\left(e^{\widehat{\widetilde{x}}_k-\whx_k}-1\right)
-\dfrac{1}{\mu}\left(e^{\widehat{\widetilde{x}}_k-\wx_k}-1\right)
+\lambda e^{x_{k+1}-\wx_k}-\mu e^{x_{k+1}-\whx_k}=0,
\end{equation}
\begin{equation}\label{eq: BT Toda S2}\tag{$S2$}
\dfrac{1}{\lambda}\left(e^{\wx_{k+1}-x_{k+1}}-1\right)
-\dfrac{1}{\mu}\left(e^{\whx_{k+1}-x_{k+1}}-1\right)
+\lambda e^{\whx_{k+1}-\widehat{\widetilde{x}}_k}
-\mu e^{\wx_{k+1}-\widehat{\widetilde{x}}_k}=0,
\end{equation}
called superposition formulas. Then the corner equations (\ref{eq: BT Toda E1})--(\ref{eq: BT Toda E12}) are satisfied, as well.
\end{theorem}
{\bf Proof.} First of all, we show that equations (\ref{eq: BT Toda S1}) and (\ref{eq: BT Toda S2}) are indeed equivalent by virtue of (\ref{eq: BT Toda E}). For this, we re-write these equations in algebraically equivalent forms:
\begin{equation}\label{eq: BT Toda S1 3leg xk+1}
\frac{\lambda-\mu}{e^{\widehat{\widetilde{x}}_k-x_{k+1}}-\lambda\mu}=\lambda e^{x_{k+1}-\wx_k}-\mu e^{x_{k+1}-\whx_k},
\end{equation}
and
\begin{equation}\label{eq: BT Toda S2 3leg xk+1}
\frac{(\lambda-\mu)e^{x_{k+1}-\widehat{\wx}_k}}{1-\lambda\mu e^{x_{k+1}-\widehat{\wx}_k}}=
\dfrac{1}{\mu}\left(e^{\whx_{k+1}-x_{k+1}}-1\right)-\dfrac{1}{\lambda}\left(e^{\wx_{k+1}-x_{k+1}}-1\right),
\end{equation}
respectively. The left-hand sides of the latter two equations are equal. Thus, their difference coincides with (\ref{eq: BT Toda E}).

 Second, we show that equations (\ref{eq: BT Toda S1}) and (\ref{eq: BT Toda S2}) yield (\ref{eq: BT Toda E1}). (For (\ref{eq: BT Toda E2}) everything is absolutely analogous.) For this aim, we re-write these equations in still other algebraically equivalent forms. Namely, (\ref{eq: BT Toda S1}) is equivalent to
\begin{equation}\label{eq: BT Toda S1 3leg wx}
e^{\widehat{\widetilde{x}}_k-\wx_k}=
 \lambda\mu e^{x_{k+1}-\wx_k}+\frac{\mu-\lambda}{\mu e^{\wx_k-\whx_k}-\lambda},
\end{equation}
while (\ref{eq: BT Toda S2}) with $k$ replaced by $k-1$ is equivalent to
\begin{equation}\label{eq: BT Toda S2 3leg wx}
\lambda\mu e^{\wx_k-\widehat{\widetilde{x}}_{k-1}}=
 e^{\wx_k-x_k}+\frac{\lambda-\mu}{\mu -\lambda e^{\whx_k-\wx_k}}.
\end{equation}
An obvious linear combination of these expressions leads to
\[
\frac{1}{\mu} e^{\widehat{\widetilde{x}}_k-\wx_k}+\mu e^{\wx_k-\widehat{\widetilde{x}}_{k-1}}=
\lambda e^{x_{k+1}-\wx_k}+\frac{1}{\lambda} e^{\wx_k-x_k}+\frac{\lambda-\mu}{\lambda\mu},
\]
which is nothing but (\ref{eq: BT Toda E1}).

Third, we observe that the sum of equations (\ref{eq: BT Toda E}), (\ref{eq: BT Toda S1}) and (\ref{eq: BT Toda S2}) is nothing but the corner equation (\ref{eq: BT Toda E12}). \endpf
\medskip

{\bf Remark.} Observe that each of the equations (\ref{eq: BT Toda S1}) and (\ref{eq: BT Toda S2}) is a quad-equation with respect to
\[
\left(e^{x_{k+1}},e^{\wx_k},e^{\whx_k},e^{\widehat{\widetilde{x}}_k}\right),\quad {\rm resp.}\quad
\left(e^{x_{k+1}},e^{\wx_{k+1}},e^{\whx_{k+1}},e^{\widehat{\widetilde{x}}_k}\right),
\]
i.e., can be formulated as vanishing of a multi-affine polynomial of the four specified variables. Equations
(\ref{eq: BT Toda S1 3leg xk+1}) and (\ref{eq: BT Toda S2 3leg xk+1}) are then interpreted as the three-leg forms of the original equations, centered at $x_{k+1}$. Similarly, equations (\ref{eq: BT Toda S1 3leg wx}) and (\ref{eq: BT Toda S2 3leg wx}) are the three-leg forms of the original equations, centered at $\wx_k$.
\medskip

\begin{figure}[tbp]
\subfloat[]{
\begin{tikzpicture}[auto,scale=2,>=stealth',inner sep=3pt]
   \node (x) at (0,0) [circle,fill,thick,label=-135:$x$,label=45:$\left(E\right)$] {};
   \node (x1) at (2,0) [circle,fill,thick,label=135:$\left(E_{1}\right)$,label=-45:$\wx$] {};
   \node (x2) at (0,2) [circle,fill,thick,label=-45:$\left(E_{2}\right)$,label=135:$\whx$] {};
   \node (x12) at (2,2) [circle,fill,thick,label=45:$\widehat{\wx}$,label=-135:$\left(E_{12}\right)$] {};
   \node (x121) at (2.25,2) [circle,fill,thick] {};
   \node (x122) at (2,2.25) [circle,fill,thick] {};
   \draw [thick,->] (x) to (x1) to (x12);
   \draw [thick,->] (x) to (x2) to (x12);
   \draw [thick,->] (x) to (x1) to (x121);
   \draw [thick,->] (x) to (x2) to (x122);
\end{tikzpicture}
}\qquad
\subfloat[]{
\begin{tikzpicture}[auto,scale=2,>=stealth',inner sep=3pt]
   \node (x) at (0,0) [circle,fill,thick,{label=-135:$\left(x,p\right)$}] {};
   \node (x1) at (2,0) [circle,fill,thick,{label=-45:$\left(\wx,\wip\right)$}] {};
   \node (x11) at (1.75,0.25) [circle,fill,thick] {};
   \node (x2) at (0,2) [circle,fill,thick,{label=135:$\left(\whx,\whp\right)$}] {};
   \node (x22) at (0.25,1.75) [circle,fill,thick] {};
   \node (x12) at (2,1.75) [circle,fill,thick,{label=45:$\big(\widehat{\wx},\widehat{\wip}\big)$}] {};
   \node (x121) at (2.25,1.5) [circle,fill,thick] {};
   \node (x122) at (1.75,2) [circle,fill,thick] {};
   \node (x123) at (1.5,2.25) [circle,fill,thick] {};
   \draw [thick,->] (x) to (x1);
   \draw [thick,->] (x) to node {$F_{\lambda}$} (x11);
   \draw [thick,->] (x) to (x2);
   \draw [thick,->] (x) to node [swap] {$F_{\mu}$} (x22);
   \draw [thick,->] (x2) to (x122);
   \draw [thick,->] (x2) to (x123);
   \draw [thick,->] (x22) to (x121);
   \draw [thick,->] (x22) to node{$F_{\lambda}$} (x12);
   \draw [thick,->] (x1) to node [swap] {$F_{\mu}$} (x122);
   \draw [thick,->] (x1) to (x121);
   \draw [thick,->] (x11) to (x123);
   \draw [thick,->] (x11) to (x12);
\end{tikzpicture}
}
\caption{Consistency of multi-time Euler-Lagrange equations in the case of periodic boundary conditions: (a) Given the fields $x$, $\wx$, $\whx$ satisfying the corner equation (\ref{eq: BT Toda E}), each of the corner equations (\ref{eq: BT Toda E1}), (\ref{eq: BT Toda E2}) has two solutions $\widehat{\wx}$. One of the values for $\widehat{\wx}$ obtained from (\ref{eq: BT Toda E1}) coincides with one of the values for $\widehat{\wx}$ obtained from (\ref{eq: BT Toda E2}). This common value of $\widehat{\wx}$, together with $\wx$ and $\whx$, satisfies (\ref{eq: BT Toda E12}). (b) The four branches of $F_\lambda\circ F_\mu$ and the four branches of $F_\mu\circ F_\lambda$ coincide pairwise.}
\label{Fig: consistency 4-valued}
\end{figure}
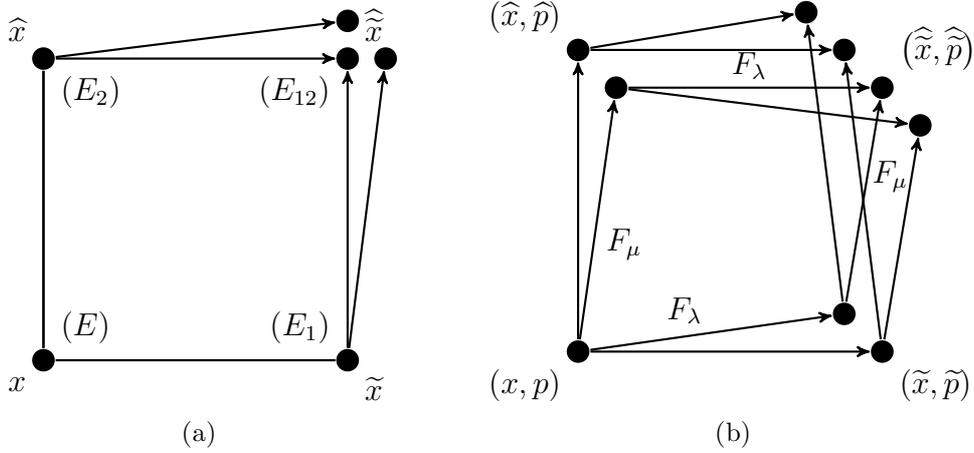

This theorem allows us to achieve an exhaustive understanding of the consistency for double-valued B\"acklund transformations. First, suppose that we are given the fields $x$, $\wx$, $\whx$ satisfying the corner equation (\ref{eq: BT Toda E}). Each of equations (\ref{eq: BT Toda E1}), (\ref{eq: BT Toda E2}) produces two values for $\widehat{\wx}$. Then consistency is reflected in the following fact: one of the values for $\widehat{\wx}$ obtained from (\ref{eq: BT Toda E1}) coincides with one of the values for $\widehat{\wx}$ obtained from (\ref{eq: BT Toda E2}), see Fig.~\ref{Fig: consistency 4-valued}(a). Indeed, this common value is nothing but $\widehat{\wx}$ obtained from the superposition formulas (\ref{eq: BT Toda S1}), (\ref{eq: BT Toda S2}), as in Theorem \ref{th: superposition}.

The ``loose ends'' on Fig.~\ref{Fig: consistency 4-valued}(a) are best explained by considering the (double-valued) maps $F_\lambda$, $F_\mu$, i.e., by working with the variables $(x,p)$ rather than with the variables $x$ alone. Indeed, each of the compositions $F_\lambda\circ F_\mu$ and $F_\mu\circ F_\lambda$ is four-valued. It follows from Theorem \ref{th: superposition} that their branches must pairwise coincide, as shown on Fig.~\ref{Fig: consistency 4-valued}(b). Indeed, Theorem \ref{th: superposition} delivers four possible values for $(\wx,\whx,\widehat{\wx})$ satisfying all corner equations (\ref{eq: BT Toda E})--(\ref{eq: BT Toda E12}), namely one $\widehat{\wx}$ for each of the four possible combinations of $(\wx,\whx)$.

\begin{theorem}\label{th: Toda closure}
The discrete multi-time Lagrangian 1-form is closed on any solution of the corner equations (\ref{eq: BT Toda E})--(\ref{eq: BT Toda E12}).
\end{theorem}
\noindent {\bf Proof.} First of all, we show that the closure relation $\ell(\lambda,\mu)=0$ is equivalent to
\begin{equation}\label{eq: BT Toda for closeness}
\sum_{k=1}^N(\widehat{\widetilde{x}}_k-\wx_k-\whx_k+x_k)=0\quad \Leftrightarrow\quad
\prod_{k=1}^N e^{\widehat{\widetilde{x}}_k-\wx_k-\whx_k+x_k}=1.
\end{equation}
This can be done in two different ways. On one hand, combining (\ref{eq: BT Toda S1}), (\ref{eq: BT Toda S2}) with (\ref{eq: BT Toda E}), we arrive at the two formulas
\begin{eqnarray}
  \dfrac{1}{\lambda}e^{\wx_{k+1}-x_{k+1}}-\dfrac{1}{\mu}e^{\whx_{k+1}-x_{k+1}}
  -\dfrac{1}{\lambda} e^{\widehat{\widetilde{x}}_k-\whx_k}
  +\dfrac{1}{\mu} e^{\widehat{\widetilde{x}}_k-\wx_k} &=& 0,
  \label{eq: BT Toda superp 1}\\
  \lambda e^{x_{k+1}-\wx_k}-\mu e^{x_{k+1}-\whx_k}
  -\lambda e^{\whx_{k+1}-\widehat{\widetilde{x}}_k}
  +\mu e^{\wx_{k+1}-\widehat{\widetilde{x}}_k} & = & 0.
  \label{eq: BT Toda superp 2}
\end{eqnarray}
By virtue of these formulas, most of the terms on the left-hand side of (\ref{eq: BT closure}) with the Lagrange function (\ref{eq: BT Toda Lagr}) cancel, leaving us with
\[
\ell(\lambda,\mu)=\left(\frac{1}{\lambda}-\frac{1}{\mu}\right)
\sum_{k=1}^N(\widehat{\widetilde{x}}_k-\wx_k-\whx_k+x_k).
\]
For an alternative proof of the fact that $\ell(\lambda,\mu)=0$ is equivalent to (\ref{eq: BT Toda for closeness}), we can refer to the spectrality criterium stating that  $\ell(\lambda,\mu)=0$ is equivalent to $\partial \Lambda(x,\wx;\lambda)/\partial \lambda$ being an integral of motion for $F_\mu$. One easily computes:
\[
\frac{\partial \Lambda(x,\wx;\lambda)}{\partial \lambda}=-\frac{1}{\lambda}\sum_{k=1}^N p_k+\frac{1}{\lambda^2}\sum_{k=1}^N(\wx_k-x_k),
\]
the first sum on the right-hand side being an obvious integral of motion.

Now the desired result (\ref{eq: BT Toda for closeness}) can be derived from the following form of the superposition formula:
\begin{equation}\label{eq: BT Toda superp}
    e^{\widehat{\widetilde{x}}_k-\wx_k-\whx_k+x_{k+1}}=\dfrac{\lambda e^{\whx_{k+1}}-\mu e^{\wx_{k+1}}}{\lambda e^{\whx_{k}}-\mu e^{\wx_{k}}},
\end{equation}
which is in fact equivalent to either of equations (\ref{eq: BT Toda superp 1}), (\ref{eq: BT Toda superp 2}). In the periodic case (\ref{eq: BT Toda for closeness}) follows directly by multiplying equations (\ref{eq: BT Toda superp}) for $1\le k\le N$, in the open-end case equation (\ref{eq: BT Toda superp}) holds true for $1\le k\le N-1$ and has to be supplemented by the boundary counterparts
\begin{equation}\label{eq: BT Toda superp open boundary}
 e^{x_{1}}=\dfrac{\lambda e^{\whx_1}-\mu e^{\wx_1}}{\lambda-\mu}, \quad
 e^{\widehat{\widetilde{x}}_N-\wx_N-\whx_N}=\dfrac{\lambda-\mu}{\lambda e^{\whx_N}-\mu e^{\wx_N}},
\end{equation}
which are equivalent to (\ref{eq: BT Toda E}) for $k=1$, resp. to (\ref{eq: BT Toda E12}) for $k=N$.
\endpf

Thus, we have proved that the quantity $\sum_{k=1}^N(\wx_k-x_k)$ is an integral of motion for  $F_\mu$ with an arbitrary $\mu$. Actually, it is not difficult to give a matrix expression for this integral. The participating matrices are nothing but transition matrices of the zero curvature representation for $F_\lambda$, but the latter notion is not necessary for establishing the result.
\begin{theorem}\label{th: BT Toda zcr}
Set
\begin{equation*}\label{eq: BT Toda zcr L}
    L_k(x,p;\lambda)=\begin{pmatrix} 1+\lambda p_k & -\lambda^2 e^{x_k} \\
                                     e^{-x_k} & 0 \end{pmatrix},
\end{equation*}
and
\begin{equation*}\label{eq: BT Toda zcr T}
    T_N(x,p;\lambda)=L_N(x,p;\lambda) \cdots L_2(x,p;\lambda)L_1(x,p;\lambda).
\end{equation*}
Then in the periodic case the quantity $\prod_{k=1}^N e^{\wx_k-x_k}$ is an eigenvalue of $T_N(x,p;\lambda)$, while in the open-end case it is equal to the $(11)$-entry of $T_N(x,p;\lambda)$.
\end{theorem}
{\bf Proof.} We use the following notation for the action of matrices from $GL(2,\mathbb C)$ on $\mathbb C$ by M\"obius transformations:
\[
\begin{pmatrix} a & b \\ c & d \end{pmatrix}[z]=\frac{az+b}{cz+d}.
\]
With this notation, we can re-write the first equation in (\ref{eq: BT1 Toda}) as
\[
e^{\wx_k}=e^{x_k}\Big((1+\lambda p_k)-\lambda^2 e^{x_k-\wx_{k-1}}\Big)=L_k(x,p;\lambda)[e^{\wx_{k-1}}].
\]
This is equivalent to saying that
\[
L_k(x,p;\lambda)\begin{pmatrix} e^{\wx_{k-1}} \\ 1 \end{pmatrix} \sim
\begin{pmatrix} e^{\wx_k} \\ 1 \end{pmatrix}.
\]
The proportionality coefficient is easily determined by comparing the second components of these vectors:
\begin{equation}\label{eq: BT Toda zcr transition}
L_k(x,p;\lambda)\begin{pmatrix} e^{\wx_{k-1}} \\ 1 \end{pmatrix} = e^{\wx_{k-1}-x_k}
\begin{pmatrix} e^{\wx_k} \\ 1 \end{pmatrix}.
\end{equation}
Now the claim in the periodic case follows immediately, with the corresponding eigenvector of $T_N(x,p;\lambda)$ being $\big(e^{\wx_N}, 1\big)^{\rm T}$. In the open-end case, equation (\ref{eq: BT Toda zcr transition}) holds true for $2\le k\le N$, and has to be supplemented by the following two relations:
\[
L_1(x,p;\lambda)\begin{pmatrix} 1 \\ 0 \end{pmatrix} = e^{-x_1}\begin{pmatrix} e^{\wx_1} \\ 1 \end{pmatrix}, \quad
{\rm and}\quad
\begin{pmatrix} 1 & 0 \end{pmatrix}\begin{pmatrix} e^{\wx_N} \\ 1 \end{pmatrix}
= e^{\wx_N}.
\]
As a consequence,
\[
\begin{pmatrix} 1 & 0 \end{pmatrix}T_N(x,p;\lambda)\begin{pmatrix} 1 \\ 0 \end{pmatrix}=\prod_{k=1}^N e^{\wx_k-x_k}. \qquad\qquad \blacksquare
\]
\smallskip

In the subsequent sections, we prove similar results for B\"acklund transformations for all the remaining Toda-type systems. For each of them, 
\begin{itemize}
\item we find superposition formulas which yield commutativity of B\"acklund transformations, both in the single-valued case of open-end boundary conditions and in the double-valued case of periodic boundary conditions;
\item we prove the spectrality property, so that the discrete 1-form $\cL$ is closed on solutions of the Euler-Lagrange equations. The proof is based on superposition formulas. This result provides also the existence of a big number of common integrals for the whole family $F_\lambda$. Actually, there are sufficiently many common integrals in involution to ensure complete integrability in the Liouville-Arnold sense;
\item and we give an expression of the corresponding conserved quantity $\partial \Lambda(x,\wx;\lambda)/\partial \lambda$ in terms of canonically conjugate variables $(x,p)$. This is done with the help of the monodromy matrix of the corresponding discrete time zero curvature representation, which is {\em derived} directly from equations of motion in an unambiguous and algorithmic way. This gives a further support to the idea pushed forward in \cite{S12}, that B\"acklund transformations can serve as zero curvature representations for themselves.
\end{itemize}

\section{B\"acklund transformations for dual Toda lattice}
\label{sect: BT dual Toda}

Our second example constitute B\"acklund transformations for the dual Toda lattice (\ref{eq: dual Toda}) which are given by equations of the type (\ref{eq: BT1}):
\begin{equation}\label{eq: BT1 dual Toda}
F_\lambda:\ \left\{\begin{array}{l}
e^{p_k}=(\wx_k-x_k)(\lambda+x_k-\wx_{k-1}), \vspace{0.1truecm}\\
e^{\wip_k}=(\wx_k-x_k)(\lambda+x_{k+1}-\wx_k).
\end{array}\right.
\end{equation}
The corresponding Lagrangian is given by
\begin{equation*}\label{eq: BT dual Toda Lagr}
\Lambda(x,\wx;\lambda)=\sum_{k=1}^N\psi(\wx_k-x_k)-\sum_{k=1}^N \psi(\lambda+x_{k+1}-\wx_k),
\end{equation*}
where $\psi(\xi)=\xi\log\xi-\xi$. The standard single-time Euler-Lagrange equations coincide with (\ref{eq: d dual Toda}) with $h=\lambda^{-1}$. Recall that for the dual Toda lattice we only consider periodic boundary conditions.

To establish commutativity of the maps $F_\lambda$, $F_\mu$, we consider the system of corner equations:
\begin{equation} \label{eq: BT dual Toda E}\tag{$E$}
 (\wx_k-x_k)(\lambda+x_k-\wx_{k-1})=(\whx_k-x_k)(\mu+x_k-\whx_{k-1}),
\end{equation}
\begin{equation} \label{eq: BT dual Toda E1}\tag{$E_1$}
 (\wx_k-x_k)(\lambda+x_{k+1}-\wx_k)=
 (\widehat{\wx}_k-\wx_k)(\mu+\wx_k-\widehat{\widetilde{x}}_{k-1}),
\end{equation}
\begin{equation} \label{eq: BT dual Toda E2}\tag{$E_2$}
 (\whx_k-x_k)(\mu+x_{k+1}-\whx_k)=
 (\widehat{\wx}_k-\whx_k)(\lambda+\whx_k-\widehat{\wx}_{k-1}),
\end{equation}
\begin{equation} \label{eq: BT dual Toda E12}\tag{$E_{12}$}
 (\widehat{\wx}_k-\whx_k)(\lambda+\whx_{k+1}-\widehat{\wx}_k)=
 (\widehat{\wx}_k-\wx_k)(\mu+\wx_{k+1}-\widehat{\wx}_k).
\end{equation}

\begin{theorem}\label{th: superposition BT dual Toda}
Suppose that the fields $x$, $\wx$, $\whx$ satisfy corner equations (\ref{eq: BT dual Toda E}). Define the fields $\widehat{\widetilde{x}}$ by any of the following two superposition formulas, which are equivalent by virtue of (\ref{eq: BT dual Toda E}):
\begin{equation}\label{eq: BT dual Toda S1}\tag{$S1$}
(\widehat{\wx}_k-\whx_k)(\lambda+x_{k+1}-\wx_k)=
(\widehat{\wx}_k-\wx_k)(\mu+x_{k+1}-\whx_k),
\end{equation}
\begin{equation}\label{eq: BT dual Toda S2}\tag{$S2$}
(\wx_{k+1}-x_{k+1})(\lambda+\whx_{k+1}-\widehat{\wx}_k)=
(\whx_{k+1}-x_{k+1})(\mu+\wx_{k+1}-\widehat{\wx}_k).
\end{equation}
Then the corner equations (\ref{eq: BT dual Toda E1})--(\ref{eq: BT dual Toda E12}) are satisfied, as well.
\end{theorem}
{\bf Proof.} Each of the equations (\ref{eq: BT dual Toda S1}) and (\ref{eq: BT dual Toda S2}) is a quad-equation with respect to
\[
\big(x_{k+1},\wx_k,\whx_k,\widehat{\widetilde{x}}_k\big),\quad {\rm resp.}\quad
\big(x_{k+1},\wx_{k+1},\whx_{k+1},\widehat{\widetilde{x}}_k\big).
\]
The three-leg forms of these equations, centered at $x_{k+1}$, are:
\begin{equation*}\label{eq: BT dual Toda S1 3leg xk+1}
\frac{\lambda+x_{k+1}-\widehat{\wx}_k}{\mu+x_{k+1}-\widehat{\wx}_k}=
\frac{\lambda+x_{k+1}-\wx_k}{\mu+x_{k+1}-\whx_k},
\end{equation*}
and
\begin{equation*}\label{eq: BT dual Toda S2 3leg xk+1}
\frac{\lambda+x_{k+1}-\widehat{\wx}_k}{\mu+x_{k+1}-\widehat{\wx}_k}=
\frac{\whx_{k+1}-x_{k+1}}{\wx_{k+1}-x_{k+1}},
\end{equation*}
respectively. Their quotient coincides with (\ref{eq: BT dual Toda E}).

The three-leg forms of equations (\ref{eq: BT dual Toda S1}) and (\ref{eq: BT dual Toda S2}), centered at $\wx_k$, are:
\begin{equation*}\label{eq: BT dual Toda S1 3leg wx}
\frac{\whx_k-\wx_k}{\lambda-\mu+\whx_k-\wx_k}=
 \frac{\widehat{\wx}_k-\wx_k}{\lambda+x_{k+1}-\wx_k},
\end{equation*}
and
\begin{equation*}\label{eq: BT dual Toda S2 3leg wx}
\frac{\whx_{k+1}-\wx_{k+1}}{\lambda-\mu+\whx_{k+1}-\wx_{k+1}}=
 \frac{\wx_{k+1}-x_{k+1}}{\mu+\wx_{k+1}-\widehat{\wx}_k},
\end{equation*}
respectively. The quotient of these two equations (the second with the downshifted index $k$) coincides with (\ref{eq: BT dual Toda E1}).

The three-leg forms of superposition formulas (\ref{eq: BT dual Toda S1}) and (\ref{eq: BT dual Toda S2}), centered at $\widehat{\wx}_k$, are:
\begin{equation*}\label{eq: BT dual Toda S1 3leg wwx}
\frac{\lambda+x_{k+1}-\widehat{\wx}_k}{\mu+x_{k+1}-\widehat{\wx}_k}=
 \frac{\widehat{\wx}_k-\wx_k}{\widehat{\wx}_k-\whx_k},
\end{equation*}
and
\begin{equation*}\label{eq: BT dual Toda S2 3leg wwx}
\frac{\lambda+x_{k+1}-\widehat{\wx}_k}{\mu+x_{k+1}-\widehat{\wx}_k}=
 \frac{\lambda+\whx_{k+1}-\widehat{\wx}_k}{\mu+\wx_{k+1}-\widehat{\wx}_k},
\end{equation*}
respectively. The quotient of these two equations coincides with (\ref{eq: BT dual Toda E12}). \endpf
\medskip

\begin{theorem}\label{th: dual Toda closure}
The discrete multi-time Lagrangian 1-form is closed on any solution of the corner equations (\ref{eq: BT dual Toda E})--(\ref{eq: BT dual Toda E12}).
\end{theorem}
\noindent {\bf Proof.} With the help of the spectrality criterium, we see that the claim of the theorem is equivalent to
\begin{equation}\label{eq: BT dual Toda for closeness}
    \prod_{k=1}^N(\lambda+\whx_{k+1}-\widehat{\wx}_k)=\prod_{k=1}^N(\lambda+x_{k+1}-\wx_k).
\end{equation}
To prove this relation, we observe that superposition formulas (\ref{eq: BT dual Toda S1}) and  (\ref{eq: BT dual Toda S2}) admit the following further equivalent formulations:
\begin{equation*}\label{eq: BT dual Toda S1 4side}
(\lambda-\mu)(\lambda+x_{k+1}-\wx_k)=
(\lambda-\mu+\whx_k-\wx_k)(\lambda+x_{k+1}-\widehat{\wx}_k),
\end{equation*}
and
\begin{equation*}\label{eq: BT dual Toda S2 4side}
(\lambda-\mu)(\lambda+\whx_{k+1}-\widehat{\wx}_k)=
(\lambda-\mu+\whx_{k+1}-\wx_{k+1})(\lambda+x_{k+1}-\widehat{\wx}_k),
\end{equation*}
respectively. There follows:
\begin{equation*}\label{eq: BT dual Toda superp}
\frac{\lambda+\whx_{k+1}-\widehat{\wx}_k}{\lambda+x_{k+1}-\wx_k}
=\frac{\mu-\lambda+\wx_{k+1}-\whx_{k+1}}{\mu-\lambda+\wx_k-\whx_k}.
\end{equation*}
Under the periodic boundary conditions, this formula yields (\ref{eq: BT dual Toda for closeness}).
\endpf

\begin{theorem}\label{th: BT dual Toda zcr}
Set
\begin{equation*}\label{eq: BT dual Toda zcr L}
    L_k(x,p;\lambda)=\begin{pmatrix} -x_k & x_k(\lambda+x_k)+e^{p_k} \\
                                     -1 & \lambda+x_k \end{pmatrix},
\end{equation*}
and
\begin{equation*}\label{eq: BT dual Toda zcr T}
    T_N(x,p;\lambda)=L_N(x,p;\lambda) \cdots L_2(x,p;\lambda)L_1(x,p;\lambda).
\end{equation*}
Then, under the periodic boundary conditions,  the quantity $\prod_{k=1}^N(\lambda+x_{k+1}-\wx_k)$ is an eigenvalue of $T_N(x,p;\lambda)$.
\end{theorem}
{\bf Proof.} We can re-write the first equation in (\ref{eq: BT1 dual Toda}) as
\[
\wx_k=x_k+\frac{e^{p_k}}{\lambda+x_k-\wx_{k-1}}=
\frac{(\lambda+x_k)x_k+e^{p_k}-x_k\wx_{k-1}}{\lambda+x_k-\wx_{k-1}}=
L_k(x,p;\lambda)[\wx_{k-1}].
\]
This is equivalent to
\[
L_k(x,p;\lambda)\begin{pmatrix} \wx_{k-1} \\ 1 \end{pmatrix} \sim
\begin{pmatrix} \wx_k \\ 1 \end{pmatrix}.
\]
The proportionality coefficient is determined by comparing the second components of these two vectors:
\begin{equation*}\label{eq: BT dual Toda zcr transition}
L_k(x,p;\lambda)\begin{pmatrix} \wx_{k-1} \\ 1 \end{pmatrix} = (\lambda+x_k-\wx_{k-1})
\begin{pmatrix} \wx_k \\ 1 \end{pmatrix}.
\end{equation*}
Now the claim follows immediately from the periodic boundary conditions, with the corresponding eigenvector of $T_N(x,p;\lambda)$ being $\big(\wx_N, 1\big)^{\rm T}$.
\endpf

\section{B\"acklund transformations for modified Toda lattice}
\label{sect: BT mod Toda}

Our third example constitute B\"acklund transformations for the modified Toda lattice (\ref{eq: mod Toda}) which are given by equations of the type (\ref{eq: BT1}):
\begin{equation}\label{eq: BT1 mod Toda}
F_\lambda:\ \left\{\begin{array}{l}
e^{p_k}=\left(e^{\wx_k-x_k}-1\right)\left(\lambda+e^{x_k-\wx_{k-1}}\right),\vspace{0.1truecm}\\
e^{\wip_k}=\left(e^{\wx_k-x_k}-1\right)\left(\lambda+e^{x_{k+1}-\wx_k}\right).
\end{array}\right.
\end{equation}
Like for the standard Toda lattice, in the open-end case the first equations in (\ref{eq: BT1 mod Toda}) are uniquely solved for $\wx_1$, $\wx_2$, $\ldots$, $\wx_N$ (in this order), and in the periodic case $\wx_k$ are double-valued functions of $(x,p)$.

The corresponding Lagrangian is given in the periodic case by
\begin{equation*}\label{eq: BT mod Toda Lagr per}
\Lambda(x,\wx;\lambda)=\sum_{k=1}^N\psi(\wx_k-x_k;-1)-\sum_{k=1}^N \psi(x_{k+1}-\wx_k;\lambda),
\end{equation*}
and in the open-end case by
\begin{equation*}\label{eq: BT mod Toda Lagr open}
\Lambda(x,\wx;\lambda)=\sum_{k=1}^N\psi(\wx_k-x_k;-1)-\sum_{k=1}^{N-1} \psi(x_{k+1}-\wx_k;\lambda)+(\wx_N-x_1)\log\lambda,
\end{equation*}
where
\[
\psi(\xi;\lambda)=\int_0^\xi\log(e^\eta+\lambda)d\eta.
\]
The standard single-time Euler-Lagrange equations are (\ref{eq: d mod Toda}) with $h=\lambda^{-1}$.

To establish commutativity of the maps $F_\lambda$, $F_\mu$, we consider the system of corner equations:
\begin{equation} \label{eq: BT mod Toda E}\tag{$E$}
 \left(e^{\wx_k-x_k}-1\right)\left(\lambda+e^{x_k-\wx_{k-1}}\right)=
 \left(e^{\whx_k-x_k}-1\right)\left(\mu+e^{x_k-\whx_{k-1}}\right),
\end{equation}
\begin{equation} \label{eq: BT mod Toda E1}\tag{$E_1$}
 \left(e^{\wx_k-x_k}-1\right)\left(\lambda+e^{x_{k+1}-\wx_k}\right)=
 \big(e^{\widehat{\wx}_k-\wx_k}-1\big)\big(\mu+e^{\wx_k-\widehat{\wx}_{k-1}}\big),
\end{equation}
\begin{equation} \label{eq: BT mod Toda E2}\tag{$E_2$}
 \big(e^{\whx_k-x_k}-1\big)\big(\mu+e^{x_{k+1}-\whx_k}\big)=
 \big(e^{\widehat{\wx}_k-\whx_k}-1\big)\big(\lambda+e^{\whx_k-\widehat{\wx}_{k-1}}\big),
\end{equation}
\begin{equation} \label{eq: BT mod Toda E12}\tag{$E_{12}$}
 \big(e^{\widehat{\wx}_k-\whx_k}-1\big)\big(\lambda+e^{\whx_{k+1}-\widehat{\wx}_k}\big)=
 \big(e^{\widehat{\wx}_k-\wx_k}-1\big)\big(\mu+e^{\wx_{k+1}-\widehat{\wx}_k}\big).
\end{equation}

\begin{theorem}\label{th: superposition BT mod Toda}
Suppose that the fields $x$, $\wx$, $\whx$ satisfy corner equations (\ref{eq: BT mod Toda E}). Define the fields $\widehat{\wx}$ by any of the following two superposition formulas, which are equivalent by virtue of (\ref{eq: BT mod Toda E}):
\begin{equation}\label{eq: BT mod Toda S1}\tag{$S1$}
\big(e^{\widehat{\wx}_k-\whx_k}-1\big)\big(\lambda+e^{x_{k+1}-\wx_k}\big)=
\big(e^{\widehat{\wx}_k-\wx_k}-1\big)\big(\mu+e^{x_{k+1}-\whx_k}\big),
\end{equation}
\begin{equation}\label{eq: BT mod Toda S2}\tag{$S2$}
\big(e^{\wx_{k+1}-x_{k+1}}-1\big)\big(\lambda+e^{\whx_{k+1}-\widehat{\wx}_k}\big)=
\big(e^{\whx_{k+1}-x_{k+1}}-1\big)\big(\mu+e^{\wx_{k+1}-\widehat{\wx}_k}\big).
\end{equation}
Then the corner equations (\ref{eq: BT mod Toda E1})--(\ref{eq: BT mod Toda E12}) are satisfied, as well.
\end{theorem}
{\bf Proof.} Each of the equations (\ref{eq: BT mod Toda S1}) and (\ref{eq: BT mod Toda S2}) is a quad-equation with respect to
\[
\big(e^{x_{k+1}},e^{\wx_k},e^{\whx_k},e^{\widehat{\wx}_k}\big),\quad {\rm resp.}\quad
\big(e^{x_{k+1}},e^{\wx_{k+1}},e^{\whx_{k+1}},e^{\widehat{\wx}_k}\big).
\]

The three-leg forms of equations (\ref{eq: BT mod Toda S1}) and (\ref{eq: BT mod Toda S2}), centered at $x_{k+1}$, are:
\begin{equation*}\label{eq: BT mod Toda S1 3leg xk+1}
\frac{\lambda+e^{x_{k+1}-\widehat{\wx}_k}}{\mu+e^{x_{k+1}-\widehat{\wx}_k}}=
\frac{\lambda+e^{x_{k+1}-\wx_k}}{\mu+e^{x_{k+1}-\whx_k}},
\end{equation*}
and
\begin{equation*}\label{eq: BT mod Toda S2 3leg xk+1}
\frac{\lambda+e^{x_{k+1}-\widehat{\wx}_k}}{\mu+e^{x_{k+1}-\widehat{\wx}_k}}=
\frac{e^{\whx_{k+1}-x_{k+1}}-1}{e^{\wx_{k+1}-x_{k+1}}-1},
\end{equation*}
respectively. Their quotient coincides with (\ref{eq: BT mod Toda E}).

The three-leg forms of equations (\ref{eq: BT mod Toda S1}) and (\ref{eq: BT mod Toda S2}), centered at $\wx_k$, are:
\begin{equation*}\label{eq: BT mod Toda S1 3leg wx}
\frac{e^{\whx_k-\wx_k}-1}{\lambda-\mu e^{\whx_k-\wx_k}}=
 \frac{e^{\widehat{\wx}_k-\wx_k}-1}{\lambda+e^{x_{k+1}-\wx_k}},
\end{equation*}
and
\begin{equation*}\label{eq: BT mod Toda S2 3leg wx}
\frac{e^{\whx_{k+1}-\wx_{k+1}}-1}{\lambda-\mu e^{\whx_{k+1}-\wx_{k+1}}}=
 \frac{e^{\wx_{k+1}-x_{k+1}}-1}{\mu+e^{\wx_{k+1}-\widehat{\wx}_k}},
\end{equation*}
respectively. The quotient of these two equations (the second with the downshifted index $k$) coincides with (\ref{eq: BT dual Toda E1}).

The three-leg forms of superposition formulas (\ref{eq: BT dual Toda S1}) and (\ref{eq: BT dual Toda S2}), centered at $\widehat{\wx}_k$, are:
\begin{equation*}\label{eq: BT mod Toda S1 3leg wwx}
\frac{\lambda+e^{x_{k+1}-\widehat{\wx}_k}}{\mu+e^{x_{k+1}-\widehat{\wx}_k}}=
 \frac{e^{\widehat{\wx}_k-\wx_k}-1}{e^{\widehat{\wx}_k-\whx_k}-1},
\end{equation*}
and
\begin{equation*}\label{eq: BT mod Toda S2 3leg wwx}
\frac{\lambda+e^{x_{k+1}-\widehat{\wx}_k}}{\mu+e^{x_{k+1}-\widehat{\wx}_k}}=
 \frac{\lambda+e^{\whx_{k+1}-\widehat{\wx}_k}}{\mu+e^{\wx_{k+1}-\widehat{\wx}_k}},
\end{equation*}
respectively. The quotient of these two equations coincides with (\ref{eq: BT dual Toda E12}). \endpf
\medskip

\begin{theorem}\label{th: mod Toda closure}
The discrete multi-time Lagrangian 1-form is closed on any solution of the corner equations (\ref{eq: BT mod Toda E})--(\ref{eq: BT mod Toda E12}).
\end{theorem}
\noindent {\bf Proof.} With the help of the spectrality criterium, we see that the claim of the theorem in the periodic case is equivalent to
\begin{equation}\label{eq: BT mod Toda for closeness}
    P(x,\wx)=P(\whx,\widehat{\wx}),
\end{equation}
where in the periodic case
\begin{equation*}\label{eq: BT mod Toda P per}
    P(x,\wx)=\prod_{k=1}^N\big(1+\lambda e^{\wx_k-x_{k+1}}\big),
\end{equation*}
while in the open-end case
\begin{equation*}\label{eq: BT mod Toda P open}
   P(x,\wx)=e^{\wx_N-x_1} \prod_{k=1}^{N-1}\big(1+\lambda e^{\wx_k-x_{k+1}}\big).
\end{equation*}
To prove this relation, we observe that superposition formulas (\ref{eq: BT mod Toda S1}) and  (\ref{eq: BT mod Toda S2}) admit the following further equivalent formulations:
\begin{equation*}\label{eq: BT mod Toda S1 4side}
(\lambda-\mu)\big(1+\lambda e^{\wx_k-x_{k+1}}\big)=
(\lambda e^{\wx_k-\whx_k}-\mu\big)\big(1+\lambda e^{\widehat{\wx}_k-x_{k+1}}\big),
\end{equation*}
and
\begin{equation*}\label{eq: BT mod Toda S2 4side}
(\lambda-\mu)\big(1+\lambda e^{\widehat{\wx}_k-\whx_{k+1}}\big)=
(\lambda e^{\wx_{k+1}-\whx_{k+1}}-\mu\big)\big(1+\lambda e^{\widehat{\wx}_k-x_{k+1}}\big),
\end{equation*}
respectively. There follows:
\begin{equation}\label{eq: BT mod Toda superp}
\frac{1+\lambda e^{\widehat{\wx}_k-\whx_{k+1}}}{1+\lambda e^{\wx_k-x_{k+1}}}
=\frac{\lambda e^{\wx_{k+1}-\whx_{k+1}}-\mu}{\lambda e^{\wx_k-\whx_k}-\mu}.
\end{equation}
In the periodic case this formula yields (\ref{eq: BT mod Toda for closeness}). In the open-end case equation (\ref{eq: BT mod Toda superp}) holds true for $1\le k\le N-1$, and has to be supplemented with
\begin{equation}\label{eq: mod Toda aux}
\frac{\lambda e^{\wx_1-\whx_1}-\mu}{\lambda-\mu}=e^{x_1-\whx_1}, \quad
\frac{\lambda-\mu}{\lambda e^{\wx_N-\whx_N}-\mu}=e^{\widehat{\wx}_N-\wx_N},
\end{equation}
which are equivalent to (\ref{eq: BT mod Toda E}) for $k=1$, resp. to (\ref{eq: BT mod Toda E12}) for $k=N$. Product of equations (\ref{eq: BT mod Toda superp}) with $1\le k\le N-1$ and equations (\ref{eq: mod Toda aux}) yields the claim of the theorem in the open-end case, as well. \endpf

\begin{theorem}\label{th: BT mod Toda zcr}
Set
\begin{equation*}\label{eq: BT mod Toda zcr L}
    L_k(x,p;\lambda)=\begin{pmatrix} \lambda+e^{p_k} & e^{x_k} \\
                                     \lambda e^{-x_k} & 1 \end{pmatrix},
\end{equation*}
and
\begin{equation*}\label{eq: BT mod Toda zcr T}
    T_N(x,p;\lambda)=L_N(x,p;\lambda) \cdots L_2(x,p;\lambda)L_1(x,p;\lambda).
\end{equation*}
Then in the periodic case the quantity $P(x,\wx)$ from (\ref{eq: BT mod Toda P per}) is an eigenvalue of $T_N(x,p;\lambda)$, while in the open-end case its counterpart from (\ref{eq: BT mod Toda P open}) is equal, up to the factor $\lambda$, to the $(11)$-entry of $T_N(x,p;\lambda)$.
\end{theorem}
{\bf Proof.} We can re-write the first equation in (\ref{eq: BT1 mod Toda}) as
\[
e^{\wx_k}=e^{x_k}\left(1+\frac{e^{p_k}}{\lambda+e^{x_k-\wx_{k-1}}}\right)=
\frac{\big(\lambda+e^{p_k}\big)e^{\wx_{k-1}}+e^{x_k}}{\lambda e^{-x_k+\wx_{k-1}}+1}=
L_k(x,p;\lambda)[e^{\wx_{k-1}}].
\]
This is equivalent to
\[
L_k(x,p;\lambda)\begin{pmatrix} e^{\wx_{k-1}} \\ 1 \end{pmatrix} \sim
\begin{pmatrix} e^{\wx_k} \\ 1 \end{pmatrix}.
\]
The proportionality coefficient is determined by comparing the second components of these two vectors:
\begin{equation*}\label{eq: BT mod Toda zcr transition}
L_k(x,p;\lambda)\begin{pmatrix} e^{\wx_{k-1}} \\ 1 \end{pmatrix} = \big(1+\lambda e^{\wx_{k-1}-x_k}\big)
\begin{pmatrix} e^{\wx_k} \\ 1 \end{pmatrix}.
\end{equation*}
Now the claim in the periodic case follows immediately, with the corresponding eigenvector of $T_N(x,p;\lambda)$ being $\big(e^{\wx_N}, 1\big)^{\rm T}$. In the open-end case, we have to use additionally the following relations:
\[
L_1(x,p;\lambda)\begin{pmatrix} 1 \\ 0 \end{pmatrix}=
\lambda e^{-x_1}\begin{pmatrix} e^{\wx_1} \\ 1 \end{pmatrix},\quad
\begin{pmatrix} 1 & 0 \end{pmatrix}\begin{pmatrix} e^{\wx_N} \\ 1 \end{pmatrix} = e^{\wx_N}.
\]
As a consequence,
\[
\begin{pmatrix} 1 & 0 \end{pmatrix} T_N(x,p;\lambda)\begin{pmatrix} 1 \\ 0 \end{pmatrix}=
\lambda e^{\wx_N-x_1}\prod_{k=1}^{N-1}\big(1+\lambda e^{\wx_k-x_{k+1}}\big). \qquad\qquad \blacksquare
\]

\section{B\"acklund transformations for symmetric rational additive Toda-type system}
\label{sect: BT add rat}

The next example constitute B\"acklund transformations for the symmetric rational additive Toda-type system (\ref{eq: sym rat add Toda}) which are given by equations of the type (\ref{eq: BT1}):
\begin{equation}\label{eq: BT1 add rat}
F_\lambda:\ \left\{\begin{array}{l}
p_k=\dfrac{\lambda}{\wx_k-x_k}+\dfrac{\lambda}{x_k-\wx_{k-1}},\vspace{0.1truecm}\\
\wip_k=\dfrac{\lambda}{\wx_k-x_k}+\dfrac{\lambda}{x_{k+1}-\wx_k}.
\end{array}\right.
\end{equation}
The corresponding Lagrangian is given by
\begin{equation}\label{eq: BT add rat Lagr}
\Lambda(x,\wx;\lambda)=\lambda\sum_{k=1}^N\log|\wx_k-x_k|-
    \lambda\sum_{k=1}^N \log|x_{k+1}-\wx_k|.
\end{equation}
The standard single-time Euler-Lagrange equations are (\ref{eq: d sym rat add Toda}) with $h=\lambda$. 
In the open-end case, all terms with $x_1-\wx_0$ and with $x_{N+1}-\wx_N$ should be omitted both from the equations of motion (\ref{eq: BT1 add rat}) and from the Lagrangian (\ref{eq: BT add rat Lagr}). As usual, in the open-end case the first equations in (\ref{eq: BT1 add rat}) are uniquely solved for $\wx_1$, $\wx_2$, $\ldots$, $\wx_N$ (in this order) in terms of $(x,p)$, while
in the periodic case all $\wx_k$ can be expressed as infinite periodic continued fractions and are, therefore, double-valued functions of $(x,p)$.

To establish commutativity of the maps $F_\lambda$, $F_\mu$, we consider the system of corner equations:
\begin{equation} \label{eq: BT add rat E}\tag{$E$}
 \dfrac{\lambda}{\wx_k-x_k}+\dfrac{\lambda}{x_k-\wx_{k-1}}=
 \dfrac{\mu}{\whx_k-x_k}+\dfrac{\mu}{x_k-\whx_{k-1}},
\end{equation}
\begin{equation} \label{eq: BT add rat E1}\tag{$E_1$}
 \dfrac{\lambda}{\wx_k-x_k}+\dfrac{\lambda}{x_{k+1}-\wx_k}=
 \dfrac{\mu}{\widehat{\wx}_k-\wx_k}+\dfrac{\mu}{\wx_k-\widehat{\widetilde{x}}_{k-1}},
\end{equation}
\begin{equation} \label{eq: BT add rat E2}\tag{$E_2$}
 \dfrac{\mu}{\whx_k-x_k}+\dfrac{\mu}{x_{k+1}-\whx_k}=
 \dfrac{\lambda}{\widehat{\wx}_k-\whx_k}+\dfrac{\lambda}{\whx_k-\widehat{\wx}_{k-1}},
\end{equation}
\begin{equation} \label{eq: BT add rat E12}\tag{$E_{12}$}
 \dfrac{\lambda}{\widehat{\wx}_k-\whx_k}+\dfrac{\lambda}{\whx_{k+1}-\widehat{\wx}_k}=
 \dfrac{\mu}{\widehat{\wx}_k-\wx_k}+\dfrac{\mu}{\wx_{k+1}-\widehat{\wx}_k}.
\end{equation}

\begin{theorem}\label{th: superposition BT add rat}
Suppose that the fields $x$, $\wx$, $\whx$ satisfy corner equations (\ref{eq: BT add rat E}). Define the fields $\widehat{\widetilde{x}}$ by any of the following two superposition formulas, which are equivalent by virtue of (\ref{eq: BT add rat E}):
\begin{equation}\label{eq: BT add rat S1}\tag{$S1$}
\mu(\widehat{\widetilde{x}}_k-\whx_k)(x_{k+1}-\wx_k)=
\lambda(\widehat{\widetilde{x}}_k-\wx_k)(x_{k+1}-\whx_k),
\end{equation}
\begin{equation}\label{eq: BT add rat S2}\tag{$S2$}
\mu(\wx_{k+1}-x_{k+1})(\whx_{k+1}-\widehat{\wx}_k)=
\lambda(\whx_{k+1}-x_{k+1})(\wx_{k+1}-\widehat{\wx}_k).
\end{equation}
Then the corner equations (\ref{eq: BT add rat E1})--(\ref{eq: BT add rat E12}) are satisfied, as well.
\end{theorem}
{\bf Proof.} Observe that equations (\ref{eq: BT add rat S1}) and (\ref{eq: BT add rat S2}) are quad-equations with respect to
\[
\big(x_{k+1},\wx_k,\whx_k,\widehat{\widetilde{x}}_k\big),\quad {\rm resp.}\quad
\big(x_{k+1},\wx_{k+1},\whx_{k+1},\widehat{\widetilde{x}}_k\big),
\]
namely the cross-ratio equations (Q1$_{\delta=0}$ in the notation of ABS list \cite{ABS}).

The three-leg forms of these equations, centered at $x_{k+1}$, are:
\begin{equation*}\label{eq: BT add rat S1 3leg xk+1}
\frac{\lambda-\mu}{\widehat{\wx}_k-x_{k+1}}=\frac{\lambda}{\wx_k-x_{k+1}}-
\frac{\mu}{\whx_k-x_{k+1}},
\end{equation*}
and
\begin{equation*}\label{eq: BT add rat S2 3leg xk+1}
\frac{\lambda-\mu}{\widehat{\wx}_k-x_{k+1}}=\frac{\lambda}{\wx_{k+1}-x_{k+1}}-
\frac{\mu}{\whx_{k+1}-x_{k+1}},
\end{equation*}
respectively. Their difference coincides with (\ref{eq: BT add rat E}).

The three-leg forms of equations (\ref{eq: BT add rat S1}) and (\ref{eq: BT add rat S2}), centered at $\wx_k$, resp. at $\wx_{k+1}$, are:
\begin{equation*}\label{eq: BT add rat S1 3leg wx}
\frac{\lambda-\mu}{\whx_k-\wx_k}=\frac{\lambda}{x_{k+1}-\wx_k}-\frac{\mu}{\widehat{\wx}_k-\wx_k},
\end{equation*}
and
\begin{equation*}\label{eq: BT add rat S2 3leg wx}
\frac{\lambda-\mu}{\whx_{k+1}-\wx_{k+1}}=\frac{\lambda}{x_{k+1}-\wx_{k+1}}-
\frac{\mu}{\widehat{\wx}_k-\wx_{k+1}},
\end{equation*}
respectively. The difference of these two equations (the second one with $k$ replaced by $k-1$) coincides with (\ref{eq: BT add rat E1}).

The three-leg forms of equations (\ref{eq: BT add rat S1}) and (\ref{eq: BT add rat S2}), centered at $\widehat{\wx}_k$, are:
\begin{equation*}\label{eq: BT add rat S1 3leg wwx}
\frac{\lambda-\mu}{x_{k+1}-\widehat{\wx}_k}=
 \frac{\lambda}{\whx_k-\widehat{\wx}_k}-\frac{\mu}{\wx_k-\widehat{\wx}_k},
\end{equation*}
and
\begin{equation*}\label{eq: BT add rat S2 3leg wwx}
\frac{\lambda-\mu}{x_{k+1}-\widehat{\wx}_k}=
 \frac{\lambda}{\whx_{k+1}-\widehat{\wx}_k}-\frac{\mu}{\wx_{k+1}-\widehat{\wx}_k},
\end{equation*}
respectively. The difference of these two equations coincides with (\ref{eq: BT add rat E12}). \endpf
\medskip

\begin{theorem}\label{th: add rat closure}
The discrete multi-time Lagrangian 1-form is closed on any solution of the corner equations (\ref{eq: BT add rat E})--(\ref{eq: BT add rat E12}).
\end{theorem}
\noindent {\bf Proof.} With the help of the spectrality criterium, we see that the claim of the theorem is equivalent to
\begin{equation}\label{eq: BT add rat for closeness}
    P(x,\wx)=P(\whx,\widehat{\wx}),
\end{equation}
where in the periodic case
\begin{equation*}\label{eq: BT add rat P}
    P(x,\wx)=\frac{\prod_{k=1}^N(x_{k+1}-\wx_k)}{\prod_{k=1}^N(\wx_k-x_k)},
\end{equation*}
while in the open-end case the product in the numerator of the latter formula is over $1\le k\le N-1$ only. To prove this relation, we re-write the cross-ratio equations (\ref{eq: BT add rat S1}), (\ref{eq: BT add rat S2}) in the following equivalent forms:
\[
(\lambda-\mu)(\widehat{\wx}_k-\whx_k)(x_{k+1}-\wx_k)=\lambda(\wx_k-\whx_k)(x_{k+1}-\widehat{\wx}_k),
\]
resp.
\[
(\lambda-\mu)(\whx_{k+1}-\widehat{\wx}_k)(\wx_{k+1}-x_{k+1})=
\lambda(\wx_{k+1}-\whx_{k+1})(x_{k+1}-\widehat{\wx}_k).
\]
As a consequence, we arrive at the following superposition formula:
\begin{equation}\label{eq: BT add rat superp}
\frac{\whx_{k+1}-\widehat{\wx}_k}{\widehat{\wx}_k-\whx_k}
=\frac{x_{k+1}-\wx_k}{\wx_{k+1}-x_{k+1}}\cdot\frac{\wx_{k+1}-\whx_{k+1}}{\wx_k-\whx_k}.
\end{equation}
In the periodic case this formula yields (\ref{eq: BT add rat for closeness}). In the open-end case equation (\ref{eq: BT add rat superp}) holds true for $1\le k\le N-1$, and has to be supplemented with
\begin{equation}\label{eq: add rat aux3}
\frac{\lambda}{\widehat{\wx}_N-\whx_N}=\frac{\lambda-\mu}{\wx_N-\whx_N}, \quad
\frac{\lambda}{\wx_1-x_1}=\frac{\lambda-\mu}{\wx_1-\whx_1}.
\end{equation}
Product of equations (\ref{eq: BT add rat superp}) with $1\le k\le N-1$ with equations (\ref{eq: add rat aux3}) yields the claim of the theorem in the open-end case, as well. \endpf

\begin{theorem}\label{th: BT add rat zcr}
Set
\begin{equation*}\label{eq: BT add rat zcr L}
    L_k(x,p;\lambda)=I+\lambda^{-1}\begin{pmatrix} p_kx_k & -p_kx_k^2 \\
                                     p_k & -p_kx_k \end{pmatrix},
\end{equation*}
and
\begin{equation*}\label{eq: BT add rat zcr T}
    T_N(x,p;\lambda)=L_N(x,p;\lambda) \cdots L_2(x,p;\lambda)L_1(x,p;\lambda).
\end{equation*}
Then in the periodic case the quantity
\[
 \frac{\prod_{k=1}^N(\wx_k-x_{k+1})}{\prod_{k=1}^N(\wx_k-x_k)}
\]
is an eigenvalue of $T_N(x,p;\lambda)$, while in the open-end case its counterpart,
\[
 \frac{\prod_{k=1}^{N-1}(\wx_k-x_{k+1})}{\prod_{k=1}^N(\wx_k-x_k)},
\]
is equal to the $(21)$-entry of $T_N(x,p;\lambda)$.
\end{theorem}
{\bf Proof.} We can re-write the first equation in (\ref{eq: BT1 add rat}) as
\[
\wx_k=x_k+\frac{\lambda}{p_k-\dfrac{\lambda}{x_k-\wx_{k-1}}}=
\frac{p_kx_k(x_k-\wx_{k-1})-\lambda \wx_{k-1}}{p_k(x_k-\wx_{k-1})-\lambda}=
L_k(x,p;\lambda)[\wx_{k-1}].
\]
This is equivalent to
\[
L_k(x,p;\lambda)\begin{pmatrix} \wx_{k-1} \\ 1 \end{pmatrix} \sim
\begin{pmatrix} \wx_k \\ 1 \end{pmatrix}.
\]
The proportionality coefficient is determined by comparing the second components of these two vectors and is equal to
\[
1-\lambda^{-1}p_k(x_k-\wx_{k-1})=1-(x_k-\wx_{k-1})\left(\frac{1}{\wx_k-x_k}+\frac{1}{x_k-\wx_{k-1}}\right)=
\frac{\wx_{k-1}-x_k}{\wx_k-x_k}.
\]
Thus,
\begin{equation}\label{eq: BT add rat zcr transition}
L_k(x,p;\lambda)\begin{pmatrix} \wx_{k-1} \\ 1 \end{pmatrix} = \frac{\wx_{k-1}-x_k}{\wx_k-x_k}
\begin{pmatrix} \wx_k \\ 1 \end{pmatrix}.
\end{equation}
Now the claim in the periodic case follows immediately, with the corresponding eigenvector of $T_N(x,p;\lambda)$ being $\big(\wx_N, 1\big)^{\rm T}$. In the open-end case, equation (\ref{eq: BT add rat zcr transition}) holds true for $2\le k\le N$, and has to be supplemented by the following two relations:
\[
L_1(x,p;\lambda)\begin{pmatrix} 1 \\ 0 \end{pmatrix} = \frac{1}{\wx_1-x_1}
\begin{pmatrix} \wx_1 \\ 1 \end{pmatrix}, \quad
{\rm and}\quad
\begin{pmatrix} 0 & 1 \end{pmatrix}\begin{pmatrix} \wx_N \\ 1 \end{pmatrix} = 1.
\]
As a consequence,
\[
\begin{pmatrix} 0 & 1 \end{pmatrix}T_N(x,p;\lambda)\begin{pmatrix} 1 \\ 0 \end{pmatrix}=
\frac{\prod_{k=1}^{N-1}(\wx_k-x_{k+1})}{\prod_{k=1}^N(\wx_k-x_k)}.\qquad\qquad \blacksquare
\]

\section{B\"acklund transformations for symmetric rational multiplicative Toda-type system}
\label{sect: BT mult rat}

The next example constitute B\"acklund transformations for the symmetric rational multiplicative Toda-type system (\ref{eq: sym rat mult Toda}) which are given by equations of the type (\ref{eq: BT1}):
\begin{equation}\label{eq: BT1 mult rat}
F_\lambda:\ \left\{\begin{array}{l}
e^{2p_k}=\dfrac{\wx_k-x_k+\lambda}{\wx_k-x_k-\lambda} \cdot
\dfrac{x_k-\wx_{k-1}+\lambda}{x_k-\wx_{k-1}-\lambda},\vspace{0.2truecm}\\
e^{2\wip_k}=\dfrac{\wx_k-x_k+\lambda}{\wx_k-x_k-\lambda} \cdot
\dfrac{x_{k+1}-\wx_{k}+\lambda}{x_{k+1}-\wx_{k}-\lambda}.
\end{array}\right.
\end{equation}
The corresponding Lagrangian is given by
\begin{equation}\label{eq: BT mult rat Lagr}
\Lambda(x,\wx;\lambda)=\sum_{k=1}^N \psi(\wx_k-x_k;\lambda)-
\sum_{k=1}^N\psi(x_{k+1}-\wx_k;\lambda),
\end{equation}
where
$$
\psi(\xi;\lambda)=\frac{1}{2}\int_{\xi-\lambda}^{\xi+\lambda}\log\eta d\eta.
$$
The standard single-time Euler-Lagrange equations are (\ref{eq: d sym rat mult Toda}) with $h=\lambda$.
In the open-end case, all terms with $x_1-\wx_0$ and with $x_{N+1}-\wx_N$ should be omitted both from the equations of motion (\ref{eq: BT1 mult rat}) and from the Lagrangian (\ref{eq: BT mult rat Lagr}).

To establish commutativity of the maps $F_\lambda$, $F_\mu$, we consider the system of corner equations:
\begin{equation} \label{eq: BT mult rat E}\tag{$E$}
\dfrac{\wx_k-x_k+\lambda}{\wx_k-x_k-\lambda} \cdot
\dfrac{x_k-\wx_{k-1}+\lambda}{x_k-\wx_{k-1}-\lambda}=
\dfrac{\whx_k-x_k+\mu}{\whx_k-x_k-\mu} \cdot
\dfrac{x_k-\whx_{k-1}+\mu}{x_k-\whx_{k-1}-\mu},
\end{equation}
\begin{equation} \label{eq: BT mult rat E1}\tag{$E_1$}
\dfrac{\wx_k-x_k+\lambda}{\wx_k-x_k-\lambda} \cdot
\dfrac{x_{k+1}-\wx_{k}+\lambda}{x_{k+1}-\wx_{k}-\lambda}=
\dfrac{\widehat{\wx}_k-\wx_k+\mu}{\widehat{\wx}_k-\wx_k-\mu} \cdot
\dfrac{\wx_k-\widehat{\wx}_{k-1}+\mu}{\wx_k-\widehat{\wx}_{k-1}-\mu},
\end{equation}
\begin{equation} \label{eq: BT mult rat E2}\tag{$E_2$}
\dfrac{\whx_k-x_k+\mu}{\whx_k-x_k-\mu} \cdot
\dfrac{x_{k+1}-\whx_{k}+\mu}{x_{k+1}-\whx_{k}-\mu}=
\dfrac{\widehat{\wx}_k-\whx_k+\lambda}{\widehat{\wx}_k-\whx_k-\lambda} \cdot
\dfrac{\whx_k-\widehat{\wx}_{k-1}+\lambda}{\whx_k-\widehat{\wx}_{k-1}-\lambda},
\end{equation}
\begin{equation} \label{eq: BT mult rat E12}\tag{$E_{12}$}
\dfrac{\widehat{\wx}_k-\whx_k+\lambda}{\widehat{\wx}_k-\whx_k-\lambda} \cdot
\dfrac{\whx_{k+1}-\widehat{\wx}_{k}+\lambda}{\whx_{k+1}-\widehat{\wx}_{k}-\lambda}=
\dfrac{\widehat{\wx}_k-\wx_k+\mu}{\widehat{\wx}_k-\wx_k-\mu} \cdot
\dfrac{\wx_{k+1}-\widehat{\wx}_{k}+\mu}{\wx_{k+1}-\widehat{\wx}_{k}-\mu}.
\end{equation}

\begin{theorem}\label{th: superposition BT mult rat}
Suppose that the fields $x$, $\wx$, $\whx$ satisfy corner equations (\ref{eq: BT add rat E}). Define the fields $\widehat{\widetilde{x}}$ by any of the following two superposition formulas, which are equivalent by virtue of (\ref{eq: BT mult rat E}):
\begin{equation}\label{eq: BT mult rat S1}\tag{$S1$}
\mu(\widehat{\widetilde{x}}_k-\whx_k)(x_{k+1}-\wx_k)-
\lambda(\widehat{\widetilde{x}}_k-\wx_k)(x_{k+1}-\whx_k)
+\lambda\mu(\lambda-\mu)=0,
\end{equation}
\begin{equation}\label{eq: BT mult rat S2}\tag{$S2$}
\mu(\wx_{k+1}-x_{k+1})(\whx_{k+1}-\widehat{\wx}_k)-
\lambda(\whx_{k+1}-x_{k+1})(\wx_{k+1}-\widehat{\wx}_k)
+\lambda\mu(\lambda-\mu)=0.
\end{equation}
Then the corner equations (\ref{eq: BT mult rat E1})--(\ref{eq: BT mult rat E12}) are satisfied, as well.
\end{theorem}

{\bf Proof.} Observe that equations (\ref{eq: BT mult rat S1}) and (\ref{eq: BT mult rat S2}) are quad-equations with respect to
\[
\big(x_{k+1},\wx_k,\whx_k,\widehat{\widetilde{x}}_k\big),\qquad {\rm resp.}\qquad
\big(x_{k+1},\wx_{k+1},\whx_{k+1},\widehat{\widetilde{x}}_k\big),
\]
namely of the type Q1$_{\delta=1}$ from the ABS list \cite{ABS}.

The three-leg forms of these equations, centered at $x_{k+1}$, are:
\begin{equation*}\label{eq: BT mult rat S1 3leg xk+1}
\frac{x_{k+1}-\wx_k+\lambda}{x_{k+1}-\wx_k-\lambda}=
\frac{x_{k+1}-\whx_k+\mu}{x_{k+1}-\whx_k-\mu}\cdot
\frac{\widehat{\widetilde{x}}_k-x_{k+1}-\lambda+\mu}{\widehat{\widetilde{x}}_k-x_{k+1}+\lambda-\mu},
\end{equation*}
and
\begin{equation*}\label{eq: BT mult rat S2 3leg xk+1}
\dfrac{\wx_{k+1}-x_{k+1}+\lambda}{\wx_{k+1}-x_{k+1}-\lambda} =
\dfrac{\whx_{k+1}-x_{k+1}+\mu}{\whx_{k+1}-x_{k+1}-\mu} \cdot
\frac{\widehat{\widetilde{x}}_k-x_{k+1}+\lambda-\mu}{\widehat{\widetilde{x}}_k-x_{k+1}-\lambda+\mu}
\end{equation*}
respectively. Their product coincides with (\ref{eq: BT mult rat E})
with $k$ replaced by $k+1$.

The three-leg forms of equations (\ref{eq: BT mult rat S1}) and (\ref{eq: BT mult rat S2}), centered at $\wx_k$, resp. at $\wx_{k+1}$, are:
\begin{equation*}\label{eq: BT mult rat S1 3leg wx}
\frac{x_{k+1}-\wx_k+\lambda}{x_{k+1}-\wx_k-\lambda}=
\frac{\widehat{\wx}_k - \wx_k+\mu}{\widehat{\wx}_k - \wx_k-\mu}\cdot
\frac{\whx_k - \wx_k +\lambda-\mu}{\whx_k - \wx_k -\lambda+\mu},
\end{equation*}
and
\begin{equation*}\label{eq: BT mult rat S2 3leg wx}
\dfrac{\wx_{k+1}-x_{k+1}+\lambda}{\wx_{k+1}-x_{k+1}-\lambda}=
\dfrac{\wx_{k+1}-\widehat{\wx}_{k}+\mu}{\wx_{k+1}-\widehat{\wx}_{k}-\mu}
\cdot
\frac{\whx_{k+1} - \wx_{k+1} -\lambda+\mu}{\whx_{k+1} - \wx_{k+1} +\lambda-\mu},
\end{equation*}
respectively. The product of these two equations (the second one with $k$ replaced by $k-1$) coincides with (\ref{eq: BT mult rat E1}).

The three-leg forms of equations (\ref{eq: BT mult rat S1}) and (\ref{eq: BT mult rat S2}), centered at $\widehat{\wx}_k$, are:
\begin{equation*}\label{eq: BT mult rat S1 3leg wwx}
\dfrac{\widehat{\wx}_{k}-\whx_{k}+\lambda}{\widehat{\wx}_{k}-\whx_{k}-\lambda}=
\dfrac{\widehat{\wx}_{k}-\wx_{k}+\mu}{\widehat{\wx}_{k}-\wx_{k}-\mu}
\cdot
\frac{\widehat{\wx}_{k} -x_{k+1} +\lambda-\mu}{\widehat{\wx}_{k} -x_{k+1} -\lambda+\mu},
\end{equation*}
and
\begin{equation*}\label{eq: BT mult rat S2 3leg wwx}
\dfrac{\whx_{k+1}-\widehat{\wx}_{k}+\lambda}{\whx_{k+1}-\widehat{\wx}_{k}-\lambda}=
\dfrac{\wx_{k+1}-\widehat{\wx}_{k}+\mu}{\wx_{k+1}-\widehat{\wx}_{k}-\mu}\cdot
\frac{\widehat{\wx}_{k} -x_{k+1} -\lambda+\mu}{\widehat{\wx}_{k} -x_{k+1} +\lambda-\mu},
\end{equation*}
respectively. The product of these two equations coincides with (\ref{eq: BT mult rat E12}). \endpf
\medskip

\begin{theorem}\label{th: mult rat closure}
The discrete multi-time Lagrangian 1-form is closed on any solution of the corner equations (\ref{eq: BT mult rat E})--(\ref{eq: BT mult rat E12}).
\end{theorem}
\noindent {\bf Proof.} Spectrality criterium requires to prove that the following quantity is an integral of motion for $F_\mu$:
\begin{equation*}\label{eq: BT mult rat P}
    P(x,\wx)=\exp(-2\partial\Lambda(x,\wx;\lambda)/\partial \lambda)=\frac{\prod_{k=1}^N\left((x_{k+1}-\wx_k)^2-\lambda^2\right)}
    {\prod_{k=1}^N\left((\wx_k-x_k)^2-\lambda^2\right)}
\end{equation*}
(in the periodic case;  in the open-end case the product in the numerator of the latter formula is over $1\le k\le N-1$ only). Taking into account the obvious integral of motion
\begin{equation*}\label{eq: BT mult rat imp}
\prod_{k=1}^N e^{2p_k}=\prod_{k=1}^N\frac{\wx_k-x_k+\lambda}{\wx_k-x_k-\lambda}\cdot
\prod_{k=1}^N \frac{x_{k+1}-\wx_k+\lambda}{x_{k+1}-\wx_k-\lambda},
\end{equation*}
we see that we have to prove the property 
\begin{equation}\label{eq: BT mult rat for closeness}
    P(x,\wx)=P(\whx,\widehat{\wx})
\end{equation}
for the quantity for either of the following two quantities:
\begin{equation}\label{eq: BT mult rat P1}
    P_1(x,\wx)=\frac{\prod_{k=1}^N\left(x_{k+1}-\wx_k+\lambda\right)}
    {\prod_{k=1}^N\left(\wx_k-x_k-\lambda\right)},
\end{equation}
\begin{equation}\label{eq: BT mult rat P2}
    P_2(x,\wx)=\frac{\prod_{k=1}^N\left(x_{k+1}-\wx_k-\lambda\right)}
    {\prod_{k=1}^N\left(\wx_k-x_k+\lambda\right)}.
\end{equation}
To prove this, we re-write superposition formulas (\ref{eq: BT mult rat S1}), (\ref{eq: BT mult rat S2}) in the following equivalent forms:
\[
(\lambda-\mu)(\widehat{\wx}_k-\whx_k \mp \lambda)(x_{k+1}-\wx_k\pm\lambda)=
\lambda(\wx_k-\whx_k\mp\lambda\pm\mu)(x_{k+1}-\widehat{\wx}_k\pm\lambda\mp\mu),
\]
and
\[
(\lambda-\mu)(\whx_{k+1}-\widehat{\wx}_k\pm\lambda)(\wx_{k+1}-x_{k+1}\mp\lambda)=
\lambda(\wx_{k+1}-\whx_{k+1}\mp\lambda\pm\mu)(x_{k+1}-\widehat{\wx}_k\pm\lambda\mp\mu).
\]
As a consequence, we arrive at the following superposition formula:
\begin{equation}\label{eq: BT mult rat superp}
\frac{\whx_{k+1}-\widehat{\wx}_k\pm\lambda}{\widehat{\wx}_k-\whx_k\mp\lambda}=
\frac{x_{k+1}-\wx_k\pm\lambda}{\wx_{k+1}-x_{k+1}\mp\lambda}
\cdot
\frac{\wx_{k+1}-\whx_{k+1}\mp\lambda\pm\mu}{\wx_k-\whx_k\mp\lambda\pm\mu}.
\end{equation}
In the periodic case, the latter formula yields (\ref{eq: BT mult rat for closeness}) for both quantities
(\ref{eq: BT mult rat P1}), (\ref{eq: BT mult rat P2}). In the open-end case equation (\ref{eq: BT mult rat superp}) holds true for $1\le k\le N-1$, and has to be supplemented with the following two relations:
\begin{equation*}\label{eq: mult rat aux3}
\frac{\lambda}{\widehat{\wx}_N-\whx_N\mp\lambda}=\frac{\lambda-\mu}{\wx_N-\whx_N\mp\lambda\pm\mu}, \qquad
\frac{\lambda}{\wx_1-x_1\mp\lambda}=\frac{\lambda-\mu}{\wx_1-\whx_1\mp\lambda\pm\mu},
\end{equation*}
which are equivalent to equation (\ref{eq: BT mult rat E12}) for $k=N$, resp. to equation (\ref{eq: BT mult rat E}) for $k=1$. \endpf

\begin{theorem}\label{th: BT mult rat zcr}
Set
\begin{equation*}\label{eq: BT mult rat zcr L}
    L_k(x,p;\lambda)=\begin{pmatrix} \lambda(e^{2p_k}+1)+x_k(e^{2p_k}-1) &
                                                  (\lambda^2-x_k^2)(e^{2p_k}-1) \\
                                     e^{2p_k}-1 &
                                                  \lambda(e^{2p_k}+1)-x_k(e^{2p_k}-1)
                      \end{pmatrix},
\end{equation*}
and
\begin{equation*}\label{eq: BT mult rat zcr T}
    T_N(x,p;\lambda)=L_N(x,p;\lambda) \cdots L_2(x,p;\lambda)L_1(x,p;\lambda).
\end{equation*}
Then in the periodic case quantity
\[
(2\lambda)^N\ \frac{\prod_{k=1}^{N}(\wx_k-x_{k+1}-\lambda)}{\prod_{k=1}^N(\wx_k-x_k-\lambda)},
\]
is an eigenvalue of $T_N(x,p;\lambda)$, while in the open-end case its counterpart,
\[
(2\lambda)^N\ \frac{\prod_{k=1}^{N-1}(\wx_k-x_{k+1}-\lambda)}{\prod_{k=1}^N(\wx_k-x_k-\lambda)},
\]
is equal to the $(21)$-entry of $T_N(x,p;\lambda)$.
\end{theorem}
{\bf Proof.} We can re-write the first equation in (\ref{eq: BT1 mult rat}) as
\begin{eqnarray}
\wx_k & = & \frac{e^{2p_k}(\lambda+x_k)(x_k-\wx_{k-1}-\lambda)+(\lambda-x_k)(x_k-\wx_{k-1}+\lambda)}
 {e^{2p_k}(x_k-\wx_{k-1}-\lambda)-(x_k-\wx_{k-1}+\lambda)}
\nonumber \\
      & = & \frac{(\lambda(e^{2p_k}+1)+x_k(e^{2p_k}-1))\wx_{k-1}+(\lambda^2-x_k^2)(e^{2p_k}-1)}
 {(e^{2p_k}-1)\wx_{k-1}+\lambda (e^{2p_k}+1)-x_k(e^{2p_k}-1)}
\nonumber \\
      & = & L_k(x,p;\lambda)[\wx_{k-1}]. \nonumber
\end{eqnarray}
This is equivalent to
\[
L_k(x,p;\lambda)\begin{pmatrix} \wx_{k-1} \\ 1 \end{pmatrix} \sim
\begin{pmatrix} \wx_k \\ 1 \end{pmatrix}.
\]
The proportionality coefficient is determined by comparing the second components of these two vectors and is equal to
\begin{eqnarray*}
 &   & (e^{2p_k}-1)(\wx_{k-1}-x_k)+\lambda(e^{2p_k}+1)\\
 & = & (x_k-\wx_{k-1}+\lambda)-e^{2p_k}(x_k-\wx_{k-1}-\lambda) =
 -2\lambda\frac{x_k-\wx_{k-1}+\lambda}{\wx_k-x_k-\lambda}.
\end{eqnarray*}
Thus,
\begin{equation}\label{eq: BT mult rat zcr transition}
L_k(x,p;\lambda)\begin{pmatrix} \wx_{k-1} \\ 1 \end{pmatrix} =
2\lambda\frac{\wx_{k-1}-x_k-\lambda}{\wx_k-x_k-\lambda}
\begin{pmatrix} \wx_k \\ 1 \end{pmatrix}.
\end{equation}
Now the claim in the periodic case follows immediately, with the corresponding eigenvector of $T_N(x,p;\lambda)$ being $\big(\wx_N, 1\big)^{\rm T}$. In the open-end case, equation (\ref{eq: BT mult rat zcr transition}) holds true for $2\le k\le N$, and has to be supplemented by the following two relations:
\[
L_1(x,p;\lambda)\begin{pmatrix} 1 \\ 0 \end{pmatrix}=
\frac{2\lambda}{\wx_1-x_1-\lambda}\begin{pmatrix} \wx_1 \\ 1 \end{pmatrix},\quad
{\rm and}\quad \begin{pmatrix} 0 & 1 \end{pmatrix}\begin{pmatrix} \wx_N \\ 1 \end{pmatrix}=1.
\]
As a consequence,
\[
\begin{pmatrix} 0 & 1 \end{pmatrix}T_N(x,p;\lambda)\begin{pmatrix} 1 \\ 0 \end{pmatrix}=
(2\lambda)^N\frac{\prod_{k=1}^{N-1}(\wx_k-x_{k+1}-\lambda)}{\prod_{k=1}^N(\wx_k-x_k-\lambda)}.
\qquad\qquad\blacksquare
\]

\section{B\"acklund transformations for symmetric hyperbolic multiplicative Toda-type system}
\label{sect: BT mult hyp}

Our last example constitutes B\"acklund transformations for the symmetric hyperbolic multiplicative Toda-type system (\ref{eq: sym hyp mult Toda}) which are given by equations of the type (\ref{eq: BT1}):
\begin{equation}\label{eq: BT1 mult hyp}
F_\lambda:\
\left\{\begin{array}{l}
e^{2p_k}=\dfrac{\sinh(\wx_k-x_k+\lambda)}{\sinh(\wx_k-x_k-\lambda)} \cdot
\dfrac{\sinh(x_k-\wx_{k-1}+\lambda)}{\sinh(x_k-\wx_{k-1}-\lambda)},
\vspace{.2truecm}\\
e^{2\wip_k}=\dfrac{\sinh(\wx_k-x_k+\lambda)}{\sinh(\wx_k-x_k-\lambda)} \cdot
\dfrac{\sinh(x_{k+1}-\wx_{k}+\lambda)}{\sinh(x_{k+1}-\wx_{k}-\lambda)}.
\end{array}\right.
\end{equation}
The corresponding Lagrangian is given by
\begin{equation}\label{eq: BT mult hyp Lagr}
\Lambda(x,\wx;\lambda)=\sum_{k=1}^N
\psi(\wx_k-x_k;\lambda) - \sum_{k=1}^N
\psi(x_{k+1}-\wx_k;\lambda),
\end{equation}
where
$$
\psi(\xi;\lambda)=\frac{1}{2}\int_{\xi -\lambda}^{\xi +\lambda}
\log \sinh (\eta) d \eta.
$$
The standard single-time Euler-Lagrange equations are (\ref{eq: d sym hyp mult Toda}) with $h=\lambda$.
In the open-end case, all terms with $x_1-\wx_0$ and with $x_{N+1}-\wx_N$ should be omitted both from the equations of motion (\ref{eq: BT1 mult hyp}) and from the Lagrangian (\ref{eq: BT mult hyp Lagr}).

To establish commutativity of the maps $F_\lambda$, $F_\mu$, we consider the system of corner equations:
\begin{equation} \label{eq: BT mult hyp E}\tag{$E$}
\dfrac{\sinh(\wx_k-x_k+\lambda)}{\sinh(\wx_k-x_k-\lambda)} \cdot
\dfrac{\sinh(x_k-\wx_{k-1}+\lambda)}{\sinh(x_k-\wx_{k-1}-\lambda)}=
\dfrac{\sinh(\whx_k-x_k+\mu)}{\sinh(\whx_k-x_k-\mu)} \cdot
\dfrac{\sinh(x_k-\whx_{k-1}+\mu)}{\sinh(x_k-\whx_{k-1}-\mu)},
\end{equation}
\begin{equation} \label{eq: BT mult hyp E1}\tag{$E_1$}
\!\!\!\!\dfrac{\sinh(\wx_k-x_k+\lambda)}{\sinh(\wx_k-x_k-\lambda)} \cdot
\dfrac{\sinh(x_{k+1}-\wx_{k}+\lambda)}{\sinh(x_{k+1}-\wx_{k}-\lambda)}=
\dfrac{\sinh(\widehat{\wx}_k-\wx_k+\mu)}{\sinh(\widehat{\wx}_k-\wx_k-\mu)} ,\cdot
\dfrac{\sinh(\wx_k-\widehat{\wx}_{k-1}+\mu)}{\sinh(\wx_k-\widehat{\wx}_{k-1}-\mu)},
\end{equation}
\begin{equation} \label{eq: BT mult hyp E2}\tag{$E_2$}
\dfrac{\sinh(\whx_k-x_k+\mu)}{\sinh(\whx_k-x_k-\mu)} \cdot
\dfrac{\sinh(x_{k+1}-\whx_{k}+\mu)}{\sinh(x_{k+1}-\whx_{k}-\mu)}=
\dfrac{\sinh(\widehat{\wx}_k-\whx_k+\lambda)}{\sinh(\widehat{\wx}_k-\whx_k-\lambda)} \cdot
\dfrac{\sinh(\whx_k-\widehat{\wx}_{k-1}+\lambda)}{\sinh(\whx_k-\widehat{\wx}_{k-1}-\lambda)},
\end{equation}
\begin{equation} \label{eq: BT mult hyp E12}\tag{$E_{12}$}
\!\!\!\!\dfrac{\sinh(\widehat{\wx}_k-\whx_k+\lambda)}{\sinh(\widehat{\wx}_k-\whx_k-\lambda)} \cdot
\dfrac{\sinh(\whx_{k+1}-\widehat{\wx}_{k}+\lambda)}{\sinh(\whx_{k+1}-\widehat{\wx}_{k}-\lambda)}=
\dfrac{\sinh(\widehat{\wx}_k-\wx_k+\mu)}{\sinh(\widehat{\wx}_k-\wx_k-\mu)} \cdot
\dfrac{\sinh(\wx_{k+1}-\widehat{\wx}_{k}+\mu)}{\sinh(\wx_{k+1}-\widehat{\wx}_{k}-\mu)}.
\end{equation}

\begin{theorem}\label{th: superposition BT mult hyp}
Suppose that the fields $x$, $\wx$, $\whx$ satisfy corner equations (\ref{eq: BT mult hyp E}). Define the fields $\widehat{\widetilde{x}}$ by any of the following two superposition formulas, which are
equivalent by virtue of (\ref{eq: BT mult hyp E}):
\[
(e^{4\lambda}- e^{4\mu})(e^{2 \whx_k }e^{2 \wx_k }+
e^{2 x_{k+1} }e^{2 \widehat{\widetilde{x}}_k })+
e^{2\mu}(1-e^{4\lambda})(e^{2 \whx_k }e^{2 \widehat{\widetilde{x}}_k }+
e^{2 x_{k+1} }e^{2 \wx_k })
\]
\begin{equation}\label{eq: BT mult hyp S1}\tag{$S1$}
\qquad \qquad + \, e^{2\lambda}(e^{4\mu}-1)(e^{2 \wx_k }e^{2 \widehat{\widetilde{x}}_k }+
e^{2 x_{k+1} }e^{2 \whx_k }) =0,
\end{equation}
\[
(e^{4\lambda}- e^{4\mu})(e^{2 \whx_{k+1} }e^{2 \wx_{k+1} }+
e^{2 x_{k+1} }e^{2 \widehat{\widetilde{x}}_k })+
e^{2\mu}(1-e^{4\lambda})(e^{2 \whx_{k+1}  }e^{2 \widehat{\widetilde{x}}_k }+
e^{2 x_{k+1} }e^{2 \wx_{k+1}  })
\]
\begin{equation}\label{eq: BT mult hyp S2}\tag{$S2$}
\qquad \qquad + \, e^{2\lambda}(e^{4\mu}-1)(e^{2 \wx_{k+1} }e^{2 \widehat{\widetilde{x}}_k }+
e^{2 x_{k+1} }e^{2 \whx_{k+1}  }) =0.
\end{equation}
Then corner equations (\ref{eq: BT mult hyp E1})--(\ref{eq: BT mult hyp E12}) are satisfied, as well.
\end{theorem}

{\bf Proof.} Observe that equations (\ref{eq: BT mult hyp S1}) and (\ref{eq: BT mult hyp S2}) are quad-equations with respect to
\[
\big(e^{2x_{k+1}},e^{2\wx_k},e^{2\whx_k},e^{2\widehat{\widetilde{x}}_k}\big),\qquad {\rm resp.}\qquad
\big(e^{2x_{k+1}},e^{2\wx_{k+1}},e^{2\whx_{k+1}},e^{2\widehat{\widetilde{x}}_k}\big),
\]
namely of the type Q3$_{\delta=0}$ from the ABS list \cite{ABS}.

The three-leg forms of these equations, centered at $x_{k+1}$, are:
\begin{equation*}\label{eq: BT mult hyp S1 3leg xk+1}
\frac{\sinh(x_{k+1}-\wx_k+\lambda)}{\sinh(x_{k+1}-\wx_k-\lambda)}=
\frac{\sinh(x_{k+1}-\whx_k+\mu)}{\sinh(x_{k+1}-\whx_k-\mu)}\cdot
\frac{\sinh(\widehat{\widetilde{x}}_k-x_{k+1}-\lambda+\mu)}
{\sinh(\widehat{\widetilde{x}}_k-x_{k+1}+\lambda-\mu)},
\end{equation*}
and
\begin{equation*}\label{eq: BT mult hyp S2 3leg xk+1}
\dfrac{\sinh(\wx_{k+1}-x_{k+1}+\lambda)}{\sinh(\wx_{k+1}-x_{k+1}-\lambda)} =
\dfrac{\sinh(\whx_{k+1}-x_{k+1}+\mu)}{\sinh(\whx_{k+1}-x_{k+1}-\mu)} \cdot
\frac{\sinh(\widehat{\widetilde{x}}_k-x_{k+1}+\lambda-\mu)}
{\sinh(\widehat{\widetilde{x}}_k-x_{k+1}-\lambda+\mu)},
\end{equation*}
respectively. Their product coincides with (\ref{eq: BT mult hyp E})
with $k$ replaced by $k+1$.

The three-leg forms of equations (\ref{eq: BT mult hyp S1}) and (\ref{eq: BT mult hyp S2}), centered at $\wx_k$, resp. at $\wx_{k+1}$, are:
\begin{equation*}\label{eq: BT mult hyp S1 3leg wx}
\frac{\sinh(x_{k+1}-\wx_k+\lambda)}{\sinh(x_{k+1}-\wx_k-\lambda)}=
\frac{\sinh(\widehat{\wx}_k - \wx_k+\mu)}{\sinh(\widehat{\wx}_k - \wx_k-\mu)}\cdot
\frac{\sinh(\whx_k - \wx_k +\lambda-\mu)}{\sinh(\whx_k - \wx_k -\lambda+\mu)},
\end{equation*}
and
\begin{equation*}\label{eq: BT mult hyp S2 3leg wx}
\dfrac{\sinh(\wx_{k+1}-x_{k+1}+\lambda)}{\sinh(\wx_{k+1}-x_{k+1}-\lambda)}=
\dfrac{\sinh(\wx_{k+1}-\widehat{\wx}_{k}+\mu)}{\sinh(\wx_{k+1}-\widehat{\wx}_{k}-\mu)}
\cdot
\frac{\sinh(\whx_{k+1} - \wx_{k+1} -\lambda+\mu)}{\sinh(\whx_{k+1} - \wx_{k+1} +\lambda-\mu)},
\end{equation*}
respectively. The product of these two equations (the second one with $k$ replaced by $k-1$) coincides with (\ref{eq: BT mult hyp E1}).

The three-leg forms of equations (\ref{eq: BT mult hyp S1}) and (\ref{eq: BT mult hyp S2}), centered at $\widehat{\wx}_k$, are:
\begin{equation*}\label{eq: BT mult hyp S1 3leg wwx}
\dfrac{\sinh(\widehat{\wx}_{k}-\whx_{k}+\lambda)}{\sinh(\widehat{\wx}_{k}-\whx_{k}-\lambda)}=
\dfrac{\sinh(\widehat{\wx}_{k}-\wx_{k}+\mu)}{\sinh(\widehat{\wx}_{k}-\wx_{k}-\mu)}
\cdot
\frac{\sinh(\widehat{\wx}_{k} -x_{k+1} +\lambda-\mu)}{\sinh(\widehat{\wx}_{k} -x_{k+1} -\lambda+\mu)},
\end{equation*}
and
\begin{equation*}\label{eq: BT mult hyp S2 3leg wwx}
\dfrac{\sinh(\whx_{k+1}-\widehat{\wx}_{k}+\lambda)}{\sinh(\whx_{k+1}-\widehat{\wx}_{k}-\lambda)}=
\dfrac{\sinh(\wx_{k+1}-\widehat{\wx}_{k}+\mu)}{\sinh(\wx_{k+1}-\widehat{\wx}_{k}-\mu)}\cdot
\frac{\sinh(\widehat{\wx}_{k} -x_{k+1} -\lambda+\mu)}{\sinh(\widehat{\wx}_{k} -x_{k+1} +\lambda-\mu)},
\end{equation*}
respectively. The product of these two equations coincides with (\ref{eq: BT mult hyp E12}). \endpf
\medskip

\begin{theorem}\label{th: add hyp closure}
The discrete multi-time Lagrangian 1-form is closed on any solution of the corner equations (\ref{eq: BT mult hyp E})--(\ref{eq: BT mult hyp E12}).
\end{theorem}
\noindent {\bf Proof.} With the help of the spectrality criterium, we see that the claim of the theorem is equivalent to
\begin{equation}\label{eq: BT mult hyp for closeness}
    P(x,\wx)=P(\whx,\widehat{\wx})
\end{equation}
for the quantity
\begin{equation*}\label{eq: BT mult hyp P}
    P(x,\wx)=\frac{\prod_{k=1}^N
   \sinh(x_{k+1}-\wx_k-\lambda)\sinh(x_{k+1}-\wx_k+\lambda)}
   {\prod_{k=1}^N\sinh(\wx_k-x_k -\lambda) \sinh(\wx_k-x_k + \lambda)},
\end{equation*}
(in the periodic case;  in the open-end case the product in the numerator of the latter formula is over $1\le k\le N-1$ only). Taking into account the obvious integral of motion
\begin{equation*}\label{eq: BT mult hyp imp}
\prod_{k=1}^N e^{2p_k}=\prod_{k=1}^N\frac{\sinh(\wx_k-x_k+\lambda)}{\sinh(\wx_k-x_k-\lambda)}\cdot
\prod_{k=1}^N \frac{\sinh(x_{k+1}-\wx_k+\lambda)}{\sinh(x_{k+1}-\wx_k-\lambda)},
\end{equation*}
we see that we have to prove the property (\ref{eq: BT mult hyp for closeness}) for either of the following two quantities:
\begin{equation}\label{eq: BT mult hyp P1}
    P_1(x,\wx)=\frac{\prod_{k=1}^N\sinh \left(x_{k+1}-\wx_k+\lambda\right)}
    {\prod_{k=1}^N\sinh \left(\wx_k-x_k-\lambda\right)},
\end{equation}
\begin{equation}\label{eq: BT mult hyp P2}
    P_2(x,\wx)=\frac{\prod_{k=1}^N\sinh \left(x_{k+1}-\wx_k-\lambda\right)}
    {\prod_{k=1}^N\sinh \left(\wx_k-x_k+\lambda\right)}.
\end{equation}
To prove this relation, we re-write superposition formulas (\ref{eq: BT mult hyp S1}), (\ref{eq: BT mult hyp S2}) in the following equivalent forms:
\[
\sinh(\widehat{\wx}_k-\whx_k \mp \lambda)\sinh(x_{k+1}-\wx_k\pm\lambda)=c\sinh(\wx_k-\whx_k\mp\lambda\pm\mu)\sinh(x_{k+1}-\widehat{\wx}_k\pm\lambda\mp\mu),
\]
resp.
\[
\sinh(\whx_{k+1}-\widehat{\wx}_k\pm\lambda)\sinh(\wx_{k+1}-x_{k+1}\mp\lambda)=c
\sinh(\wx_{k+1}-\whx_{k+1}\mp\lambda\pm\mu)\sinh(x_{k+1}-\widehat{\wx}_k\pm\lambda\mp\mu),
\]
where $c=(e^{4\lambda}-1)/(e^{4\lambda}-e^{4\mu})$.
As a consequence, we arrive at the following superposition formula:
\begin{equation}
\label{eq: BT mult hyp superp}
\frac{\sinh(\whx_{k+1}-\widehat{\wx}_k\pm\lambda)}{\sinh(\widehat{\wx}_k-\whx_k \mp \lambda)}=
\frac{\sinh(x_{k+1}-\wx_k\pm\lambda)}{\sinh(\wx_{k+1}-x_{k+1}\mp\lambda)} \cdot \frac{\sinh(\wx_{k+1}-\whx_{k+1}\mp\lambda\pm\mu)}{\sinh(\wx_k-\whx_k\mp\lambda\pm\mu)}.
\end{equation}

In the periodic case, the latter formula yields (\ref{eq: BT mult hyp for closeness}) for both quantities
(\ref{eq: BT mult hyp P1}), (\ref{eq: BT mult hyp P2}). In the open-end case, equation (\ref{eq: BT mult hyp superp}) holds true for $1\le k\le N-1$, and has to be supplemented with the following two relations:
\begin{equation*}\label{eq: mult hyp aux3 1}
\frac{c}{\sinh(\widehat{\wx}_N-\whx_N\mp\lambda)}=\frac{1}{\sinh(\wx_N-\whx_N\mp\lambda\pm\mu)},
\end{equation*}
\begin{equation*}\label{eq: mult hyp aux3 2}
\frac{c}{\sinh(\wx_1-x_1\mp\lambda)}=\frac{1}{\sinh(\wx_1-\whx_1\mp\lambda\pm\mu)},
\end{equation*}
which are equivalent to equation (\ref{eq: BT mult hyp E12}) for $k=N$, resp. to equation (\ref{eq: BT mult hyp E}) for $k=1$. \endpf

\begin{theorem}\label{th: BT mult hyp zcr}
Set
\begin{equation*}\label{eq: BT mult hyp zcr L}
    L_k(x,p;\lambda)=\begin{pmatrix}
    e^{4\lambda} e^{2p_k}-1
    & e^{2\lambda}e^{2 x_k } (1-e^{2p_k}) \\
      e^{2\lambda}e^{-2 x_k } (e^{2p_k}-1) &
      e^{4\lambda} - e^{2p_k}
      \end{pmatrix},
\end{equation*}
and
\begin{equation*}\label{eq: BT mult hyp zcr T}
    T_N(x,p;\lambda)=L_N(x,p;\lambda) \cdots L_2(x,p;\lambda)L_1(x,p;\lambda).
\end{equation*}
Then in the periodic case the quantity
\[
(1-e^{4\lambda})^N\ \frac{\prod_{k=1}^N
\sinh(x_{k+1}-\wx_k+\lambda)}
   {\prod_{k=1}^N\sinh(\wx_k-x_k -\lambda) },
\]
is an eigenvalue of $T_N(x,p;\lambda)$, while in the open-end case its counterpart,
\[
(1-e^{4\lambda})^N\ \frac{\prod_{k=1}^{N-1}
 \sinh(x_{k+1}-\wx_k+\lambda)}
   {\prod_{k=1}^N\sinh(\wx_k-x_k -\lambda) },
\]
is equal to the $(21)$-entry of $T_N(x,p;\lambda)$.
\end{theorem}
{\bf Proof.}
We can re-write the first equation in (\ref{eq: BT1 mult hyp}) as
$$
e^{2\wx_k}=e^{2x_k}
\frac{(e^{4\lambda}e^{2p_k}-1) e^{2\wx_{k-1} }
-e^{2\lambda}e^{2x_k} (1-e^{2p_k})}
{e^{2\lambda}(e^{2p_k}-1)e^{2\wx_{k-1} }
+e^{2x_k} (e^{4\lambda}-e^{2p_k})}
=L_k(x,p;\lambda)[e^{2\wx_{k-1} }]. \nonumber
$$
This is equivalent to
\[
L_k(x,p;\lambda)\begin{pmatrix} e^{2\wx_{k-1} } \\ 1 \end{pmatrix} \sim
\begin{pmatrix} e^{2\wx_{k} } \\ 1 \end{pmatrix}.
\]

The proportionality coefficient is determined by comparing the second components of these two vectors and is equal to
$$
e^{2\lambda}(e^{2p_k}-1)e^{2 \wx_{k-1}-2 x_k}+e^{4\lambda} - e^{2p_k}=
(1-e^{4\lambda})\frac{\sinh(x_k - \wx_{k-1}+\lambda)}{\sinh(\wx_k - x_{k}-\lambda)}.
$$
Thus,
\begin{equation}\label{eq: BT mult hyp zcr transition}
L_k(x,p;\lambda)\begin{pmatrix} e^{2\wx_{k-1}} \\ 1 \end{pmatrix} =
(1-e^{4\lambda})\frac{\sinh(x_k - \wx_{k-1}+\lambda)}{\sinh(\wx_k - x_{k}-\lambda)}
\begin{pmatrix} e^{2\wx_k} \\ 1 \end{pmatrix}.
\end{equation}
Now the claim in the periodic case follows immediately, with the corresponding eigenvector of $T_N(x,p;\lambda)$ being $\big(e^{2\wx_N}, 1\big)^{\rm T}$. In the open-end case, equation (\ref{eq: BT mult hyp zcr transition}) holds true for $2\le k\le N$, and has to be supplemented by the following two relations:
\[
L_1(x,p;\lambda)\begin{pmatrix} 1 \\ 0 \end{pmatrix}=
\frac{1-e^{4\lambda}}{\sinh(\wx_1-x_1-\lambda)}\begin{pmatrix} e^{2\wx_1} \\ 1 \end{pmatrix},\quad
{\rm and}\quad \begin{pmatrix} 0 & 1 \end{pmatrix}\begin{pmatrix} e^{2\wx_N} \\ 1 \end{pmatrix}=1.
\]
As a consequence,
\[
\begin{pmatrix} 0 & 1 \end{pmatrix}T_N(x,p;\lambda)\begin{pmatrix} 1 \\ 0 \end{pmatrix}=
(1-e^{4\lambda})^N\ \frac{\prod_{k=1}^{N-1}
 \sinh(x_{k+1}-\wx_k+\lambda)}
   {\prod_{k=1}^N\sinh(\wx_k-x_k -\lambda) }.
\qquad\qquad\blacksquare
\]

\section{Conclusions}

In a forthcoming paper, we will present results on the multi-time Lagrangian one-forms for a more general class of B\"acklund transformations, namely for systems of the relativistic Toda type. This will give us an opportunity to present an alternative approach to this theory, namely an approach from the point of view of two-dimensional integrable systems. Indeed, it is well known since \cite{A} that discrete time relativistic Toda systems are best interpreted as systems on the regular triangular lattice. A general theory of Toda-type systems on graphs and their relation to quad-graph equations has been developed in \cite{BS1}, \cite{AS}, \cite{BollSuris1}. A blend of both approaches, one- and two-dimensional, turns out to be fruitful for both ones.

Another point we plan to investigate is the quantum counterpart of the results presented here. It is well known that B\"acklund transformations for the standard Toda lattice admit a natural quantum analog, the Baxter's Q-operator \cite{PG}. The Lagrangian of the B\"acklund transformation is a quasi-classical limit of the kernel of the integral Q-operator. The spectrality property of the B\"acklund transformation is the quasi-classical limit of the Baxter's equation relating the monodromy matrix and the Q-operator \cite{Skl}. At the same time, the quantum counterpart of the whole multi-time Lagrangian theory, in particular, of the closure relation, is not yet clear. Also here, the two-dimensional point of view will be fruitful, as indicated by the work \cite{BMS} treating a solvable model of statistical mechanics, for which the thermodynamical limit of the partition function is nothing but the action functional of a certain discrete Toda-type model. In this framework, the closure relation obtains its interpretation as the thermodynamical limit of the $Z$-invariance of the partition function. A blend of both approaches will likely deliver a rather universal and simple picture.
\medskip

This research is supported by the DFG Collaborative Research Center TRR 109 ``Discretization in Geometry and Dynamics''.


\end{document}